\documentclass[twocolumn]{aastex63}

\newcommand\hd{HD\,32297}

\received{...}
\revised{...}
\accepted{...}
\submitjournal{AJ}

\shorttitle{The GPI view of the \hd\ debris disk}
\shortauthors{Duch\^ene et al.}

\begin{document}

\title{The Gemini Planet Imager view of the \hd\ debris disk}

\correspondingauthor{Gaspard Duch\^ene}
\email{gduchene@berkeley.edu}

\author[0000-0002-5092-6464]{Gaspard Duch\^ene}
\affiliation{Astronomy Department, University of California, Berkeley, CA 94720, USA}
\affiliation{Universit\'e Grenoble Alpes / CNRS, Institut de Plan\'etologie et d'Astrophysique de Grenoble, 38000 Grenoble, France}

\author[0000-0002-7670-670X]{Malena Rice}
\affiliation{Department of Astronomy, Yale University, New Haven, CT 06511, USA}

\author[0000-0001-9994-2142]{Justin Hom}
\affiliation{School of Earth and Space Exploration, Arizona State University, PO Box 871404, Tempe, AZ 85287, USA}

\author[0000-0001-9828-1848]{Joseph Zalesky}
\affiliation{School of Earth and Space Exploration, Arizona State University, PO Box 871404, Tempe, AZ 85287, USA}

\author[0000-0002-0792-3719]{Thomas M. Esposito}
\affiliation{Astronomy Department, University of California, Berkeley, CA 94720, USA}

\author[0000-0001-6205-9233]{Maxwell A. Millar-Blanchaer}
\affiliation{NASA Jet Propulsion Laboratory, California Institute of Technology, Pasadena, CA 91109, USA}

\author[0000-0003-1698-9696]{Bin Ren}
\affiliation{Department of Physics and Astronomy, Johns Hopkins University, Baltimore, MD 21218, USA}
\affiliation{Space Telescope Science Institute, Baltimore, MD 21218, USA}
\affiliation{Department of Astronomy, California Institute of Technology, 1216 East California Boulevard, Pasadena, CA 91125, USA}

\author[0000-0002-6221-5360]{Paul Kalas}
\affiliation{Astronomy Department, University of California, Berkeley, CA 94720, USA}
\affiliation{SETI Institute, Carl Sagan Center, 189 Bernardo Ave.,  Mountain View CA 94043, USA}
\affiliation{Institute of Astrophysics, FORTH, GR-71110 Heraklion, Greece}

\author[0000-0002-0176-8973]{Michael P. Fitzgerald}
\affiliation{Department of Physics \& Astronomy, 430 Portola Plaza, University of California, Los Angeles, CA 90095, USA}


\author[0000-0001-6364-2834]{Pauline Arriaga}
\affiliation{Department of Physics \& Astronomy, 430 Portola Plaza, University of California, Los Angeles, CA 90095, USA}

\author{Sebastian Bruzzone}
\affiliation{Department of Physics and Astronomy, Centre for Planetary Science and Exploration, The University of Western Ontario, London, ON N6A 3K7, Canada}

\author{Joanna Bulger}
\affiliation{Pan-STARRS Observatory, Institute for Astronomy, University of Hawai’i, 2680 Woodlawn Drive, Honolulu, HI 96822, USA}

\author[0000-0002-8382-0447]{Christine H. Chen}
\affiliation{Space Telescope Science Institute, Baltimore, MD 21218, USA}

\author{Eugene Chiang}
\affiliation{Astronomy Department, University of California, Berkeley, CA 94720, USA}
\affiliation{Earth and Planetary Science Department, University of California, Berkeley, CA 94720, USA}

\author{Tara Cotten}
\affiliation{Physics and Astronomy, University of Georgia, 240 Physics, Athens, GA 30602, USA}

\author[0000-0002-1483-8811]{Ian Czekala}
\altaffiliation{NASA Hubble Fellowship Program Sagan Fellow}
\affiliation{Astronomy Department, University of California, Berkeley, CA 94720, USA}

\author[0000-0002-4918-0247]{Robert J. De Rosa}
\affiliation{Kavli Institute for Particle Astrophysics and Cosmology, Stanford University, Stanford, CA 94305, USA}

\author[0000-0001-9290-7846]{Ruobing Dong}
\affiliation{University of Victoria, 3800 Finnerty Rd, Victoria, BC, V8P 5C2, Canada}

\author{Zachary H. Draper}
\affiliation{University of Victoria, 3800 Finnerty Rd, Victoria, BC, V8P 5C2, Canada}
\affiliation{National Research Council of Canada Herzberg, 5071 West Saanich Road, Victoria, BC V9E 2E7, Canada}

\author[0000-0002-7821-0695]{Katherine B. Follette}
\affiliation{Physics and Astronomy Department, Amherst College, 21 Merrill Science Drive, Amherst, MA 01002, USA}

\author{James R. Graham}
\affiliation{Astronomy Department, University of California, Berkeley, CA 94720, USA}

\author{Li-Wei Hung}
\affiliation{Department of Physics \& Astronomy, 430 Portola Plaza, University of California, Los Angeles, CA 90095, USA}

\author{Ronald Lopez}
\affiliation{Department of Physics \& Astronomy, 430 Portola Plaza, University of California, Los Angeles, CA 90095, USA}

\author[0000-0003-1212-7538]{Bruce Macintosh}
\affiliation{Kavli Institute for Particle Astrophysics and Cosmology, Stanford University, Stanford, CA 94305, USA}

\author{Brenda C. Matthews}
\affiliation{National Research Council of Canada Herzberg, 5071 West Saanich Road, Victoria, BC V9E 2E7, Canada}
\affiliation{University of Victoria, 3800 Finnerty Rd, Victoria, BC, V8P 5C2, Canada}

\author[0000-0002-9133-3091]{Johan Mazoyer}
\affiliation{LESIA, Observatoire de Paris, Université PSL, CNRS, Sorbonne Université, Université Paris Diderot, Sorbonne Paris Cité, 5 place Jules Janssen, 92195 Meudon, France}

\author[0000-0003-3050-8203]{Stan Metchev}
\affiliation{Department of Physics and Astronomy, Centre for Planetary Science and Exploration, The University of Western Ontario, London, ON N6A 3K7, Canada}
\affiliation{Department of Physics and Astronomy, Stony Brook University, Stony Brook, NY 11794-3800, USA}

\author{Jennifer Patience}
\affiliation{School of Earth and Space Exploration, Arizona State University, PO Box 871404, Tempe, AZ 85287, USA}

\author[0000-0002-3191-8151]{Marshall D. Perrin}
\affiliation{Space Telescope Science Institute, Baltimore, MD 21218, USA}

\author[0000-0003-0029-0258]{Julien Rameau}
\affiliation{Institut de Recherche sur les Exoplan{\`e}tes, D{\'e}partement de Physique, Universit{\'e} de Montr{\'e}al, Montr{\'e}al, QC, H3C 3J7, Canada}

\author[0000-0002-5815-7372]{Inseok Song}
\affiliation{Physics and Astronomy, University of Georgia, 240 Physics, Athens, GA 30602, USA}

\author{Kevin Stahl}
\affiliation{Department of Physics \& Astronomy, 430 Portola Plaza, University of California, Los Angeles, CA 90095, USA}

\author[0000-0003-0774-6502]{Jason Wang}
\affiliation{Department of Astronomy, California Institute of Technology, 1216 East California Boulevard, Pasadena, CA 91125, USA}
\affiliation{Astronomy Department, University of California, Berkeley, CA 94720, USA}

\author[0000-0002-9977-8255]{Schuyler Wolff}
\affiliation{Leiden Observatory, Leiden University, P.O. Box 9513, 2300 RA Leiden, The Netherlands}

\author{Ben Zuckerman}
\affiliation{Department of Physics \& Astronomy, 430 Portola Plaza, University of California, Los Angeles, CA 90095, USA}

\author[0000-0001-5172-7902]{S. Mark Ammons}
\affiliation{Lawrence Livermore National Laboratory, 7000 East Ave, Livermore, CA 94550, USA}

\author[0000-0002-5407-2806]{Vanessa P. Bailey}
\affiliation{NASA Jet Propulsion Laboratory, California Institute of Technology, Pasadena, CA 91109, USA}

\author[0000-0002-7129-3002]{Travis Barman}
\affiliation{Lunar and Planetary Laboratory, University of Arizona, Tucson AZ 85721, USA}

\author[0000-0001-6305-7272]{Jeffrey Chilcote}
\affiliation{Kavli Institute for Particle Astrophysics and Cosmology, Stanford University, Stanford, CA 94305, USA}
\affiliation{Department of Physics, University of Notre Dame, 225 Nieuwland Science Hall, Notre Dame, IN, 46556, USA}

\author{Rene Doyon}
\affiliation{Institut de Recherche sur les Exoplan{\`e}tes, D{\'e}partement de Physique, Universit{\'e} de Montr{\'e}al, Montr{\'e}al QC, H3C 3J7, Canada}

\author[0000-0003-3978-9195]{Benjamin L. Gerard}
\affiliation{University of Victoria, Department of Physics and Astronomy, 3800 Finnerty Rd, Victoria, BC V8P 5C2, Canada}
\affiliation{National Research Council of Canada Herzberg, 5071 West Saanich Rd, Victoria, BC, V9E 2E7, Canada}

\author[0000-0002-4144-5116]{Stephen J. Goodsell}
\affiliation{Gemini Observatory, 670 N. A'ohoku Place, Hilo, HI 96720, USA}

\author[0000-0002-7162-8036]{Alexandra Z. Greenbaum}
\affiliation{Department of Astronomy, University of Michigan, Ann Arbor, MI 48109, USA}

\author[0000-0003-3726-5494]{Pascale Hibon}
\affiliation{Gemini Observatory, Casilla 603, La Serena, Chile}

\author{Patrick Ingraham}
\affiliation{Large Synoptic Survey Telescope, 950N Cherry Ave., Tucson, AZ 85719, USA}

\author[0000-0002-9936-6285]{Quinn Konopacky}
\affiliation{Center for Astrophysics and Space Science, University of California San Diego, La Jolla, CA 92093, USA}

\author{J\'er\^ome Maire}
\affiliation{Center for Astrophysics and Space Science, University of California San Diego, La Jolla, CA 92093, USA}

\author[0000-0001-7016-7277]{Franck Marchis}
\affiliation{SETI Institute, Carl Sagan Center, 189 Bernardo Ave.,  Mountain View CA 94043, USA}

\author[0000-0002-5251-2943]{Mark S. Marley}
\affiliation{Space Science Division, NASA Ames Research Center, Mail Stop 245-3, Moffett Field CA 94035, USA}

\author[0000-0002-4164-4182]{Christian Marois}
\affiliation{National Research Council of Canada Herzberg, 5071 West Saanich Rd, Victoria, BC, V9E 2E7, Canada}
\affiliation{University of Victoria, 3800 Finnerty Rd, Victoria, BC, V8P 5C2, Canada}

\author[0000-0001-6975-9056]{Eric L. Nielsen}
\affiliation{SETI Institute, Carl Sagan Center, 189 Bernardo Ave.,  Mountain View CA 94043, USA}
\affiliation{Kavli Institute for Particle Astrophysics and Cosmology, Stanford University, Stanford, CA 94305, USA}

\author[0000-0001-7130-7681]{Rebecca Oppenheimer}
\affiliation{Department of Astrophysics, American Museum of Natural History, New York, NY 10024, USA}

\author{David Palmer}
\affiliation{Lawrence Livermore National Laboratory, 7000 East Ave, Livermore, CA 94550, USA}

\author{Lisa Poyneer}
\affiliation{Lawrence Livermore National Laboratory, 7000 East Ave, Livermore, CA 94550, USA}

\author{Laurent Pueyo}
\affiliation{Space Telescope Science Institute, Baltimore, MD 21218, USA}

\author[0000-0002-9246-5467]{Abhijith Rajan}
\affiliation{Space Telescope Science Institute, 3700 San Martin Drive, Baltimore, MD 21218, USA}

\author[0000-0002-9667-2244]{Fredrik T. Rantakyr\"o}
\affiliation{Gemini Observatory, Casilla 603, La Serena, Chile}

\author[0000-0003-2233-4821]{Jean-Baptiste Ruffio}
\affiliation{Kavli Institute for Particle Astrophysics and Cosmology, Stanford University, Stanford, CA, 94305, USA}

\author[0000-0002-8711-7206]{Dmitry Savransky}
\affiliation{Sibley School of Mechanical and Aerospace Engineering, Cornell University, Ithaca, NY 14853, USA}

\author{Adam C. Schneider}
\affiliation{School of Earth and Space Exploration, Arizona State University, PO Box 871404, Tempe, AZ 85287, USA}

\author[0000-0003-1251-4124]{Anand Sivaramakrishnan}
\affiliation{Space Telescope Science Institute, Baltimore, MD 21218, USA}

\author[0000-0003-2753-2819]{R\'{e}mi Soummer}
\affiliation{Space Telescope Science Institute, Baltimore, MD 21218, USA}

\author[0000-0002-9121-3436]{Sandrine Thomas}
\affiliation{Large Synoptic Survey Telescope, 950N Cherry Ave., Tucson, AZ 85719, USA}




\begin{abstract}
We present new $H$-band scattered light images of the \hd\ edge-on debris disk obtained with the Gemini Planet Imager (GPI). The disk is detected in total and polarized intensity down to a projected angular separation of 0\farcs15, or 20\,au. On the other hand, the large scale swept-back halo remains undetected, likely a consequence of its markedly blue color relative to the parent body belt. We analyze the curvature of the disk spine and estimate a radius of $\approx$100\,au for the parent body belt, smaller than past scattered light studies but consistent with thermal emission maps of the system. We employ three different flux-preserving post-processing methods to suppress the residual starlight and evaluate the surface brightness and polarization profile along the disk spine. Unlike past studies of the system, our high fidelity images reveal the disk to be highly symmetric and devoid of morphological and surface brightness perturbations. We find the dust scattering properties of the system to be consistent with those observed in other debris disks, with the exception of HR\,4796. Finally, we find no direct evidence for the presence of a planetary-mass object in the system.
\end{abstract}



\section{Introduction} \label{sec:intro}

Debris disks represent a late stage in planetary system evolution after most of the gaseous component of the protoplanetary disk has dissipated. Remnant planetesimals are thought to collide and continuously replenish these disks with small dust grains \citep{wyatt2008evolution}. Debris disks are characterized by low integrated fractional luminosity ($\tau_\mathrm{IR} = L_\mathrm{IR}/L_\mathrm{bol} \lesssim 0.01$), indicating that these are generally optically thin. While challenging, imaging these disks in scattered light in the optical and/or near-infrared often reveals offsets, asymmetries, and other irregularities, that provide a unique lens to study mature planetary systems. This is best illustrated by the $\beta$\,Pic system, the first debris disk ever imaged in which a gas giant planet responsible for a noticeable disk warp was subsequently discovered \citep{smith84betapic, burrows95, lagrange2009probable}. To date, over three dozen debris disks have been imaged in scattered light, although image fidelity is often limited by artefacts introduced by the necessary suppression of the remaining glare of the central star \citep{hughes2018debris}. 

\hd\ is a young \citep[$\leq$30\,Myr][]{kalas2005first}, A6 star\footnote{The oft-quoted A0 spectral for \hd, which can be traced back to the Henry Draper catalog, has been conclusively shown to be too hot; the best-fitting effective temperature for the stars is in the 7600--8000\,K \citep{fitzgerald2007ring, rodigas2014does}.} located 133\,pc away from the Sun\footnote{All physical lengths quoted in this paper are based on this distance, which is significantly larger than the {\it Hipparcos} distance used in previous studies \citep{perryman97}.} \citep{brown2018gaia}. It has one of the largest infrared excesses observed among main sequence stars \cite[$\tau_\mathrm{IR} \gtrsim 3 \times 10^{-3}$,][]{silverstone00} and, as a result, it is one of the best studied debris disk systems to date. In particular, it has been spatially resolved in scattered light from the optical to 4\,$\mu$m \citep[e.g.,][]{schneider2005discovery, kalas2005first, rodigas2014does}, as well as in thermal emission in the mid-infrared \citep{moerchen200712, fitzgerald2007ring} and at millimeter wavelengths \citep{maness2008carma, macgregor18}. No planet has been detected in the system, down to sensitivities of $\approx\,$2--5$\,M_\mathrm{Jup}$ \citep{bhowmik19}. In addition to a copious amount of dust, the \hd\ disk is remarkable because of the detection of Na I absorption \citep[with 5 times the column density observed in $\beta$\,Pic,][]{redfield2007gas} as well as atomic and molecular gas emission \citep{donaldson2013modeling, greaves2016, macgregor18, cataldi19}. While the number of gas detections in debris disks is steadily rising \citep{hughes2018debris}, the \hd\ system stands out as one of the most prominent such systems. The origin of this gas is still debated, but it is likely released during collisions between planetesimals, possibly very recently \citep{kral2017, cataldi19}.

Resolved images of the \hd\ debris disk revealed two spatially distinct components: a parent body belt and an extended outer halo. The halo, which was the first component detected in scattered light \citep{kalas2005first}, extends to at least 1800\,au \citep{schneider2014probing} and displays an unusually curved morphology that may be indicative of interaction with the interstellar medium \citep{debes2009interstellar}, with an undetected planet \citep{lee2016} or with the gas component of the disk \cite{lin19}, or of a recent collision in the disk as proposed by \cite{mazoyer2014hd} to explain a similar structure in the HD\,15115 disk. Either way, the halo is thought to be populated by the smallest dust grains produced by collisions in the parent belt and that are subsequently placed in high-eccentricity orbits through radiative forces.

The parent body belt, which is seen nearly exactly edge-on, has a radius of about 110--130\,au in scattered light \citep[e.g.,][]{boccaletti2012morphology, esposito2013modeling, bhowmik19}. Images are consistent with a sharp-edged inner cavity inside of this radius, while the surface density drops smoothly outwards to form the halo. This belt radius coincides with the value derived from thermal emission maps \citep{moerchen200712}, although the superior sensitivity of ALMA recently showed that the belt is radially extended and that the halo also contributes to the millimeter emission \citep{macgregor18}. Several lateral asymmetries and substructures have been proposed in scattered light images of the main belt \citep{currie2012keck, asensio2016polarimetry}. These studies are generally hampered by the necessity to employ aggressive point spread function (PSF) subtraction methods that often introduce spurious features, however, and the reality of these features remains to be firmly established \citep[e.g.,][]{milli12}.

Many of the studies discussed above have attempted to reproduce observations of the \hd\ disk to infer its dust properties. In part because each study considers different datasets -- scattered light images, thermal emission maps, entire spectral energy distribution, -- no consensus has been reached regarding the minimum grain size in the parent body belt. It could be sub-micron \citep{fitzgerald2007ring, esposito2013modeling, bhowmik19}, thus likely smaller than the blow-out size, or as large as several microns, albeit possibly with high porosity \citep{donaldson2013modeling, rodigas2014does}. The only firmly established conclusion is that the dust is strongly forward scattering, both in the optical and the near-infrared. The composition of the dust is equally contentious, ranging from a rather standard mixture of astrophysical material to pure water ice. In principle, the recent measurement of the scattered light polarization fraction in the system \citep{asensio2016polarimetry} should help reduce ambiguities, but the quality of this dataset was too low to warrant detailed modeling.
 
Here we present new scattered light observations of the central ($<$250\,au) regions of the \hd\ debris disk using the polarimetric mode of the high-contrast Gemini Planet Imager \citep[GPI,][]{macintosh2014}. We present high-fidelity scattered light images of the parent body belt in both total and polarized intensity. This allows us to assess the belt's overall geometry and to empirically characterize its dust scattering properties (\S\,\ref{sec:analysis}). We then use these quantities to constrain properties of the dust contained in the belt \S\,\ref{sec:model}. In \S\,\ref{sec:discus}, we discuss the implications of our findings before concluding in \S\,\ref{sec:conclu}.

\section{Observations and Data Reduction} \label{sec:obs}

On 2014 December 18 (UT), we observed {\hd} with GPI's polarimetric mode in the $H$ band with a 0\farcs24-diameter occulting mask. We obtained thirty-eight 60\,s frames with a half-wave plate cycling through position angles 0\degr, 22\fdg5, 45\degr\ and 67\fdg5. The observations were acquired at an airmass of 1.27 and through the target's transit, resulting in a total field rotation of 19\degr. Conditions were somewhat poorer than average, with seeing estimates of 1\farcs17 and 0\farcs82 from the Gemini Differential Image Motion Monitor and the Multi-Aperture Scintillation Sensor, respectively. Telemetry from the AO system \citep{poyneer14, bailey16} reported post-correction wavefront residuals of 150--160\,nm.

\begin{figure*}
\epsscale{1.17}
\plotone{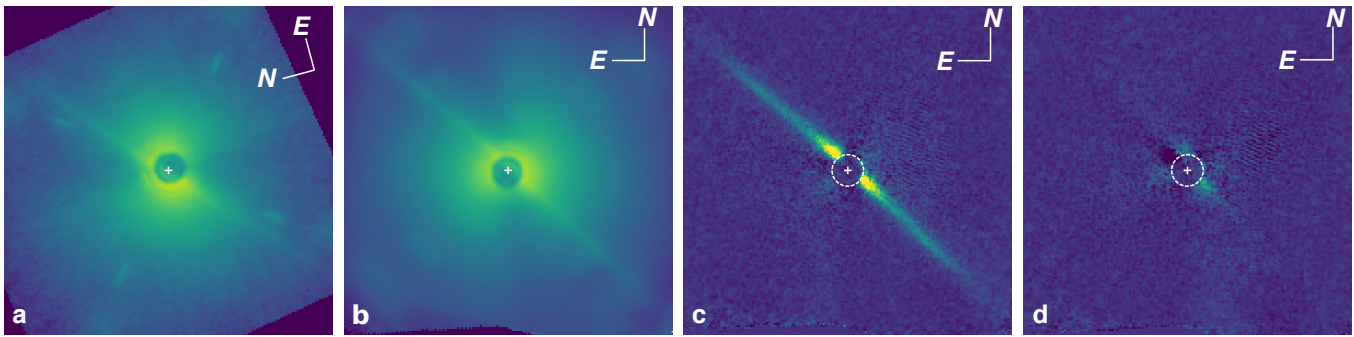}
\caption{GPI $H$-band total intensity images of \hd. A single-frame and the complete sequence total intensity images are shown on the same logarithmic stretch in panels a and b, respectively. The two right-hand side panels present the Stokes $Q_\phi$ (c) and $U_\phi$ (d) polarized intensity images, respectively, with both shown on the same linear stretch from -5 to 30 times the background noise (0.2\,mJy/arcsec$^2$). Each panel is 2\farcs5 on a side and a white plus symbol indicates the location of the star. The size of the focal plane mask is indicated by a dashed circle in panels c and d. Panels b, c and d are shown with the same orientation, while panel a is shown with the orientation of that particular frame. The reference compass rose segments have length 0\farcs25. \label{fig:raw_imgs}}
\end{figure*}

The data were processed using the GPI Data Reduction Pipeline v1.3 \citep{maire12, perrin14}. The raw data were dark subtracted, flat-fielded, cleaned of correlated detector noise, bad pixel corrected, flexure corrected, and combined into a polarization datacube (where the third dimension holds two orthogonal polarization states). Each datacube was then corrected for non-common path errors via a double differencing algorithm \citep{perrin15}. The star location was determined from the satellite spots using a radon-transform- based algorithm \citep{wang14}. The instrumental polarization was estimated by measuring the apparent stellar polarization in each polarization datacube as the mean normalized difference of pixels within 20 pixels from the star’s location. The estimated instrumental polarization was then subtracted from each pixel, scaled by the pixel’s total intensity \citep{mmb15}. While the region used to estimate the instrumental polarization includes some signal from the disk itself, only a small fraction of all pixels are affected by it and, out to that radius, the residual starlight is brighter than the disk itself. We thus estimate that this does not lead to a significant bias. The datacubes were then smoothed with a Gaussian kernel (FWHM of 1 pixel), rotated to a common orientation and combined into a Stokes datacubes via singular value decomposition \citep{perrin15}. Finally, the $[I,Q,U,V]$ Stokes cube was converted to the $[I,Q_\phi,U_\phi,V]$ ``radial Stokes" cube \citep{schmid06}, with the convention that positive $Q_\phi$ indicates a polarization vector that is perpendicular to the line joining a given point in the image to the star location, while $U_\phi$ represents polarization vectors oriented at 45\degr\ from this line.

The data were flux calibrated by measuring the brightness of the reference satellite spots \citep[][Esposito et al., submitted]{hung15}. The \hd\ disk overlaps with two of the four spots in some images, introducing a potential for a biased calibration. We therefore estimated the ADU-to-Jy conversion factors using the latter ten frames of the sequence, in which all satellite spots are cleanly separated from the disk, and we assumed that the same factors applied to the first half of the sequence. From the scatter across datacubes, the flux calibration factor is measured with a 5\% uncertainty.

\section{Observational Results} \label{sec:analysis}

\subsection{Raw Images} \label{subsec:rawimgs}

The \hd\ disk is bright enough to be detected in raw individual frames, as illustrated in Figure\,\ref{fig:raw_imgs}a. In the combined Stokes $I$ image (Figure\,\ref{fig:raw_imgs}b), the disk is strongly detected above the background of the PSF halo outside of $\approx0\farcs3$, although measuring accurate surface brightness still requires an additional step of PSF subtraction; this is performed in \S\,\ref{subsec:psfsub}. 

Because light from the star is intrinsically unpolarized, there is no leftover halo in the Stokes $Q_\phi$ and $U_\phi$ images. In the former, the disk is strongly detected from just outside the edge of the coronagraphic mask ($\approx0\farcs15$) out to a sensitivity-limited distance of about 1\farcs2 from the star. This dataset provides the smallest stellocentric distance at which the disk is clearly detected to date. Under the assumption of single scattering as in the optically thin regime, $U_\phi$ should be null throughout the image \citep{canovas15}. This is true outside of 0\farcs25, where we use the $U_\phi$ map to evaluate the noise associated with the $Q_\phi$ map by measuring the standard deviation in concentric annuli. In the inner region, however, a $U_\phi$ signal is observed at approximately the same location as the disk at a level of 5--10\% of the $Q_\phi$ signal. This could either be a consequence of multiple scattering, implying that the disk is not quite optically thin, or an indication of uncorrected polarization systematics. Because the strongest signal in the $U_\phi$ map is offset by about 2 pixels perpendicular to the disk major axis from the strongest $Q_\phi$ signal, we deem the latter interpretation as likely correct. Despite various attempts to improve data reduction, we could not find a satisfactory method to fully remove this artifact. We thus evaluate the uncertainty associated with the $Q_\phi$ map with the same method at these inner regions as at larger radii, noting that this may introduce a bias because the dispersion between pixels within annuli is not driven by random noise.

\begin{figure*}
\epsscale{1.17}
\plotone{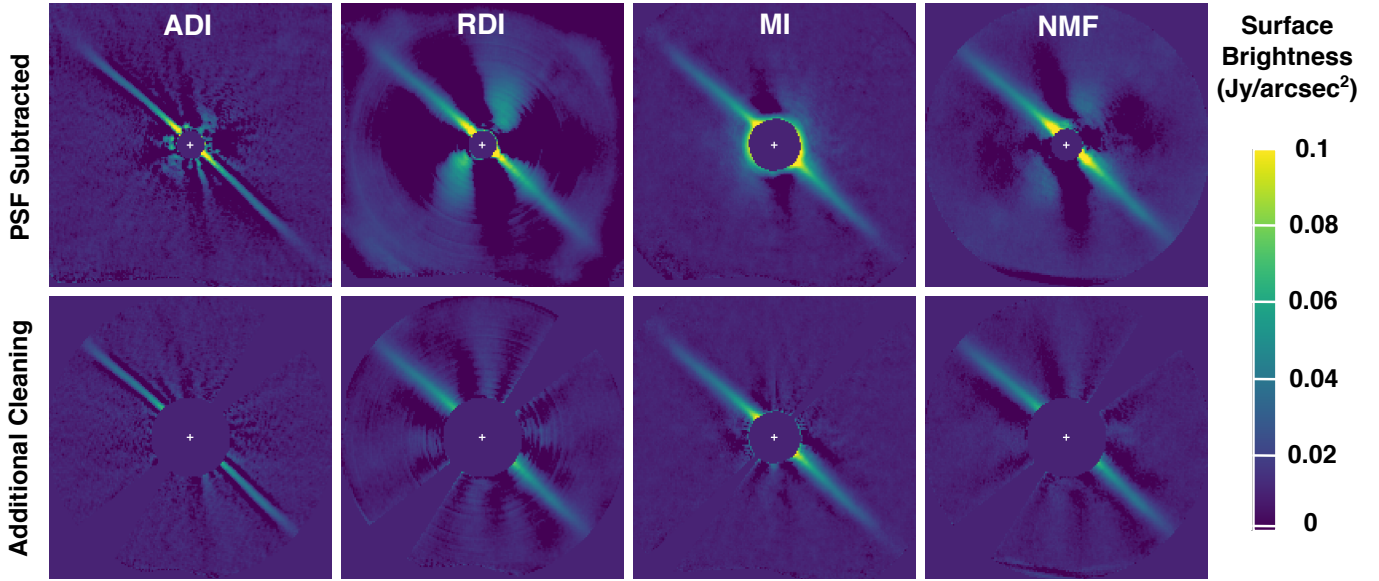}
\caption{GPI $H$-band total intensity images of the \hd\ disk after PSF subtraction, using four different methods. From left to right, the PSF subtraction methods are a conservative ADI-based pyKLIP that only uses images from the target's sequence, an RDI-based implementation of pyKLIP using other GPI $H$-band images to evaluate the PSF, a frame-by-frame MI process, and an NMF-based implementation of pyKLIP. The top row presents the product of each of these processes whereas the bottom row are our final products, after a polynomial fit is performed azimuthally and subtracted to further reduce the background. All images are shown on the same square root stretch from -0.001 to 0.1\,Jy/arcsec$^2$, except for the ADI images where the surface brightness has been multiplied by a factor of 2 to qualitatively offset self-subtraction. All images have a 2\farcs5 field of view and are oriented North up and East to the left. Numerical masks have been applied in regions with excessive subtraction residuals. \label{fig:psfsub}}
\end{figure*}

\subsection{Total Intensity Image: PSF Subtraction} \label{subsec:psfsub}

To more clearly reveal the \hd\ disk in total intensity, it is necessary to subtract the residual starlight in the Stokes $I$ image. As in previous GPIES disk analyses \citep[e.g.,][]{kalas15,draper16}, we implemented several independent methods, each with their own advantages and limitations. The resulting PSF-subtracted images are presented in the top row of Figure\,\ref{fig:psfsub}. 

First, we used a standard Angular Differential Imaging (ADI) approach with {\tt pyKLIP}-ADI \citep{wang15pyklip}, a custom implementation of the KLIP algorithm \citep{soummer12}. This method is highly effective for point source discovery but results in systematic self-subtraction of extended objects such as disks. In the particular situation of edge-on disks, strong negative ``wings" are imprinted on each side of the disk, especially when the total field rotation is modest as is the case here. To minimize self-subtraction, we adopted a conservative set of parameters, using only 5 KL modes and averaging images with 3 to 9 annuli. In the resulting image, the disk is traced all the way to the coronagraphic mask, with an apparently smooth brightness profile. 

To mitigate self-subtraction, we also used {\tt pyKLIP} with Reference Differential Imaging (RDI). Here, we first assemble a library of nearly 25,000 $H$-band images of stars observed with GPI, from which frames with known astrophysical signal or instrumental issues were removed. We then select the 500 most highly correlated with each individual frame of \hd. The PSF is then estimated by applying the same KLIP process as above to this set of reference images. While this approach prevents self-subtraction, {\tt pyKLIP}-RDI can still suffer from over-subtraction, as any astrophysical signal can be misinterpreted as a PSF "feature" by the algorithm. This is particularly relevant in the case of a bright disk like \hd, where the ratio of disk-to-PSF signal approaches or even exceeds unity in some parts of the image. We thus employed a conservative set of parameters (5 KL modes, averaged over 3-9 annuli, 500 reference PSFs chosen). Despite significant, low-frequency background fluctuations in the resulting image, the disk is clearly detected at all radii outside of the coronagraphic mask. 

To sidestep self- and over-subtraction in a different way, we also employed the mask-and-interpolate (MI) PSF subtraction at the single-frame level \citep{perrin15}. We first mask out a 15-pixel-high box centered on the disk, as well as the four satellite spots. The masked pixels are then replaced with the result of interpolating through the neighboring unmasked pixels with a fourth order polynomial function. The resulting image is then smoothed with a 13-pixel ($\approx$0\farcs18) running median box to only model the low spatial frequency component of the PSF, and it is subsequently subtracted from the original frame. Residual fluctuations in the background are significantly lower than in the RDI case, except close to the inner working angle where the interpolation scheme fails to reproduce the sharp intensity gradients of the PSF. The region interior of $\approx0$\farcs25 from the star is too uncertain to consider in our subsequent analysis, but the disk is strongly detected outside of this radius. 

Finally, we applied the Non-negative Matrix Factorization (NMF) method as implemented within {\tt pyKLIP}. NMF is an iterative method based on the decomposition of the PSF into separate components that only contain positive pixels \citep{ren18}. Similar to the RDI process, we selected the 500 most correlated frames in the library of GPI images and used the first 5 modes computed by NMF to subtract the PSF. The resulting total intensity image for \hd\ reveals a smooth brightness profile, albeit with leftover background fluctuations that are intermediate in strength between the RDI and MI methods. Like the ADI and RDI methods, the NMF method yields a strong detection of the disk all the way to the edge of the coronagraphic mask. 

Apart from the bright disk, all four PSF subtracted images are marked by a diagonal negative residual pattern (along position angles $\sim$15\degr and $\sim$190\degr). This likely is a consequence of the ``butterfly" structure of the PSF visible in the raw total intensity images (see Figure\,\ref{fig:raw_imgs}) and that is imparted by winds in the atmosphere \citep{madurowicz19}. To improve the quality of the final images, we perform a fourth-order polynomial fit in concentric annuli after masking out a vertical box centered on the disk; to improve the fit, the process handles each side of the disk separately. Effectively, this performs a second mask-and-interpolate subtraction, on a single annuli basis. The resulting images are shown in Figure\,\ref{fig:psfsub}. In the case of the RDI and NMF methods, which are characterized by more structured residuals, the subtraction residuals and amplitude of the background fluctuations become too high to produce a clean image of the disk inside of 0\farcs3 from the star. For these images, we do not attempt to measure the absolute brightness of the disk closer in. 

\subsection{Disk Morphology and Geometry} \label{subsec:morph}

\begin{figure}
\epsscale{1.17}
\plotone{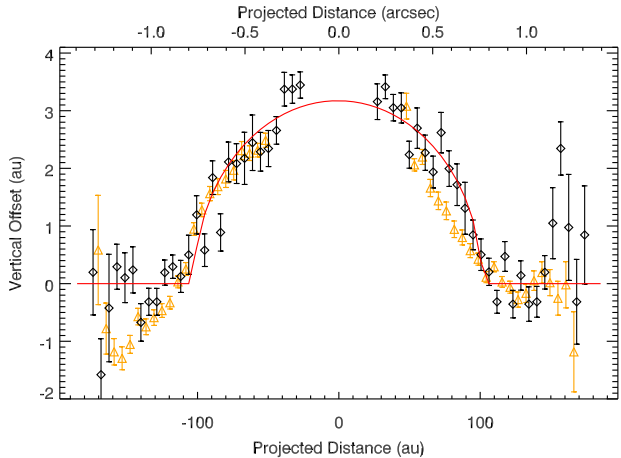}
\caption{Vertical offset between the spine of the \hd\ disk and a line at PA 47\fdg9 passing through the central star. Black diamonds and orange triangles represent estimates based on the polarized intensity and RDI total intensity images, respectively. The latter is representative of all four PSF subtraction methods employed here. The red curve is the inclined ring model that best fit the spine location in the polarized intensity image.\label{fig:vert_offset}}
\end{figure}

In both total and polarized intensity, the \hd\ disk is revealed as a sharp, almost linear feature on each side of the star along position angle (PA, measured in the usual East of North convention) of 47\fdg90$\pm$0\fdg17, as measured from the geometric fit presented below, where the uncertainty incorporates the astrometric calibration precision \citep{derosa19}. As was found in past scattered light images of the system \citep[e.g.,][]{boccaletti2012morphology}, the GPI data reveal that the spine of the disk is not perfectly straight as would be the case for a perfectly edge-on viewing geometry. Instead, the spine is slightly curved and passes to the NW of the star (see Figure\,\ref{fig:vert_offset}), indicating that this side is the front side of the disk under the assumption that scattering is preferentially in forward direction. We find no conclusive evidence of the back side of the disk.

Comparing the PSF-subtracted images of the disk to radiative transfer models can yield simultaneous constraints on both the disk geometry and dust scattering, and thus physical, properties. This is a computationally intensive task and results are often fraught with model-dependent biases and ambiguities, however. To take advantage of the high fidelity GPI images, we instead we adopt a two-step empirical approach. In the first step, we ignore the surface brightness profile along the disk, which is dictated by the surface density and scattering phase function, and focus on the spine morphology to assess the disk geometry. Having established the system geometry, we can then constrain the dust scattering properties. We defer to \S\,\ref{sec:model} the interpretation in terms of physical properties of the dust.

The marked curvature of the spine and the uniform vertical FWHM along the disk spine (see below) are best explained if the disk is radially narrow since a broad ring would yield a smeared appearance due to line of sight project effects. This allows us to employ a simple model consisting of a circular ring of radius $R_d$, whose center can be offset by $\delta_x$ from the star along the major axis, and observed with an inclination $i$. We do not explore the possibility of an offset along the minor-axis of the disk as the nearly edge-on configuration of the system renders this effect negligible. To incorporate the halo of blown-out dust to this simple model, we assume that the spine extends horizontally outside the ring ansae, i.e., with no offset from the disk major axis. On the larger scale, the halo is markedly curved, but this effect is only significant outside of the GPI field-of-view. Implicitly, this model assumes that the disk is an intrinsically narrow ring whose eccentricity is small. For a given PA of the disk major axis and ($x_\star, y_\star$) position of the star, we measure the spine by rotating the image so that the disk major axis is horizontal, binning the image by a factor of 3 (i.e., a resolution element) along the horizontal axis, and fitting a Gaussian function to the intensity profile perpendicular to the disk. Uncertainties are assigned at the pixel level based on the standard deviation in concentric 1-pixel-wide annuli and propagated through the Gaussian fit for both the total intensity maps, thus neglecting residual correlated noise. 

\begin{figure}
\epsscale{1.17}
\plotone{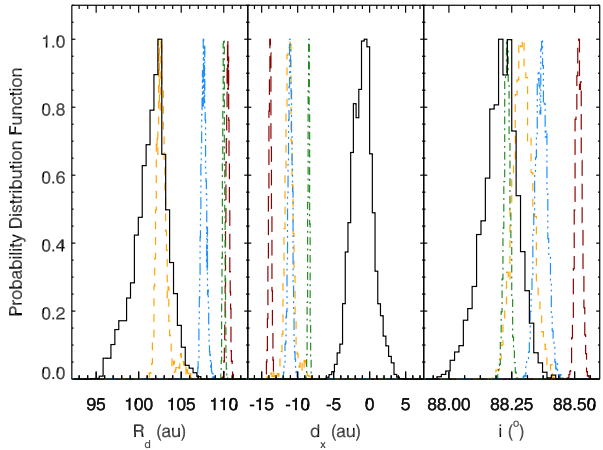}
\caption{Posterior distributions for the radius, the offset between the ring center and the star, and the inclination of the \hd\ disk. The solid black histogram represents the fit to the spine as traced in the polarized intensity image, whereas the color histograms are associated to the various PSF subtraction methods used in obtaining the total intensity image (RDI: dashed orange; ADI: dot-dashed green; MI: long-dashed red; NMF: triple-dot-dashed blue). \label{fig:geom_posteriors}}
\end{figure}

To explore the 6-dimension parameter space, we use a Metropolis-Hastings MCMC algorithm. We first perform the fit using the $Q_\phi$ (hereafter polarized intensity) image since 1) it provides a clear detection down to a smaller inner working angle, and 2) it is not subject to systematic biases introduced by PSF subtraction. As illustrated in Figure\,\ref{fig:vert_offset}, the data are reasonably well fit by this simple model ($\chi^2_{\mathrm red}=2.1$). The resulting model parameters are: $i=88$\fdg21$^{+0.06}_{-0.08}$, $R_d=101.7^{+1.5}_{-2.1}$\,au and $\delta_x=-0.9^{+1.3}_{-1.5}$\,au. The corresponding posteriors are shown in Figure\,\ref{fig:geom_posteriors}. We find no significant offset between the star and ring center, with a 3$\sigma$ upper limit on the ring eccentricity of $e < 0.05$, yielding further support to our simple geometric model. 

We then applied the same fitting method to each of the four PSF-subtracted images. The resulting posteriors are also shown in Figure\,\ref{fig:geom_posteriors}. For each dataset, the posteriors are much narrower than the posteriors from the fit to the polarized intensity image. The total intensity posteriors are also inconsistent with one another. The narrow posteriors are a consequence of the fact that uncertainties are underestimated due to correlated residuals in the PSF subtracted images. The offsets between the various posteriors is likely a consequence of subtle, but significant, modifications to the disk spine introduced by the PSF subtraction process. To illustrate this point, we show in Figure\,\ref{fig:vert_offset} the spine vertical offset observed in the RDI total intensity image assuming the exact same disk geometric parameters as the best fit to the polarized intensity image. Despite modest deviations from the spine location derived from the polarized intensity image, the fit is much worse ($\chi^2_{\mathrm red}=9.9$) and marginal differences observed on both sides (especially around positions -160 and +70\,au) conspire to push the fit towards significant eccentricity in the ring. Given this experience, we adopt the geometrical parameters obtained from fitting the polarized intensity image. 

Overall, while our geometric modeling is in reasonable agreement with past scattered light studies \citep{boccaletti2012morphology, currie2012keck, esposito2013modeling, bhowmik19}, we find a significantly smaller disk radius of $\approx100$\,au instead of $\approx130$\,au. Most of these studies used total intensity images to assess the ring geometry, thus possibly introducing a systematic bias compared to our analysis of the polarized intensity image of the disk. However, \cite{bhowmik19} also analyzed polarized observations and also favor a larger disk radius. Inspection of their Figure\,3 reveals a similar shape for the disk spine as we find here but with a global vertical displacement that can significantly bias the model fitting. This highlights the difficulty in assessing the location of the disk ansae in the edge-on configuration. We defer a more thorough discussion of the disk's viewing geometry to Section\,\ref{sec:discus}.

\begin{figure}
\epsscale{1.17}
\plotone{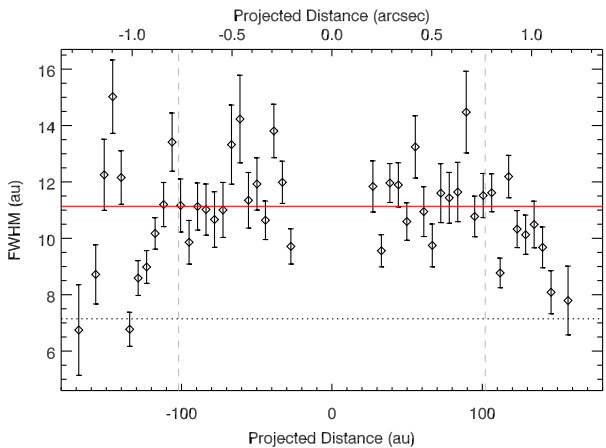}
\caption{FWHM of the \hd\ disk in the direction perpendicular to the disk midplane, as measured in the polarized intensity image of the system. The solid red line marks the weighted average over all datapoints located between the ring ansae (indicated by the vertical dashed lines) whereas the horizontal dotted line represents the intrinsic FWHM of our GPI $H$ band observations.  \label{fig:vert_fwhm}}
\end{figure}

From the same Gaussian fit as described above, we also measured the vertical FWHM of the disk. The results for the polarized intensity image are shown in Figure\,\ref{fig:vert_fwhm}. After subtracting quadratically the instrumental FWHM from the weighted average over all positions along the disk, we estimate the true FWHM of the disk to be about 0\farcs063, or 8.3\,au. While this is generally consistent with past studies \citep{boccaletti2012morphology,currie2012keck,esposito2013modeling}, we differ from these studies in that we find no significant trend as a function of stellocentric distance. We believe that the trends suggested in past analyses were affected by significant PSF subtraction artifacts. This is further supported by the fact that the FWHM measured with the RDI, MI and NMF total intensity images show a significant decline in FWHM inside of 0\farcs5, well below the value measured in the polarized intensity image. The lack of a stellocentric dependency of the disk FWHM is consistent with the hypothesis of a radially narrow disk as projection effects would result in an increase in FWHM close to the minor axis otherwise. 

\subsection{Disk Surface Brightness and Polarization Profiles} \label{subsec:surfbright}

Except for the ADI method, we have tuned our PSF subtraction methods with an eye towards preservation of the disk surface brightness profile. One of the main motivations to do this was to measure the polarization fraction in the disk. In Appendix\,\ref{subsec:recovery}, we show that injecting a model disk into an empty dataset and applying the RDI, MI and NMF methods yield surface brightness profiles that match the injected one to within 10\% or better when considering the peak surface brightness along the spine, where PSF subtraction artefacts are smallest. We then proceed and measure the surface brightness profile of the \hd\ disk using the same Gaussian as used in our geometric analysis. We also note that, since the disk is indeed an optically thin, narrow ring seen almost perfectly edge-on, limb brightening will significantly affect the observed surface brightness close to the ansae. On the other hand, because both total and polarized intensity are affected in a similar way, we expect the polarization fraction map to be mostly free of this effect. Either way, we will take the effect into account in the radiative transfer modeling presented in \S\,\ref{sec:model}.

\begin{figure}
\epsscale{1.17}
\plotone{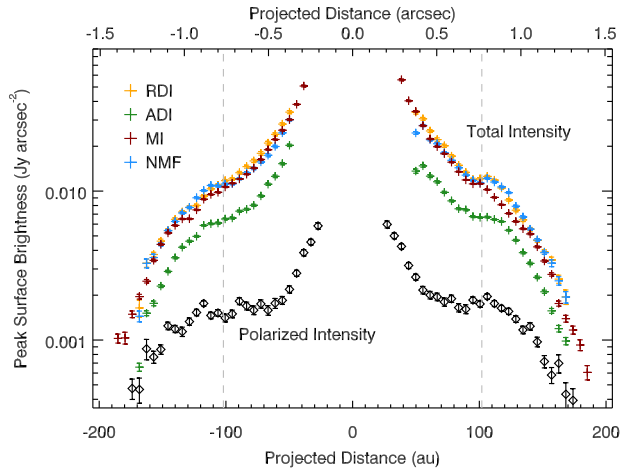}
\caption{$H$-band surface brightness profiles of the \hd\ disk in polarized intensity (black diamonds) and total intensity (colored symbols, corresponding of the different PSF subtraction methods). The vertical dashed lines indicate the disk radius as derived from the geometric fit to the disk spine.\label{fig:sb_prof}}
\end{figure}

Figure\,\ref{fig:sb_prof} presents the surface brightness profile in both polarized and total intensity. In total intensity, the profiles measured in the RDI, MI and NMF-processed images agree within $\approx$10\% of another, with the exception of a possible local maximum at $\approx$0\farcs9 in the MI image (most noticeable on the SW side of the disk). Given the amplitude of differences between the various PSF subtraction methods \citep[and in line with the surface brightness profile obtained by][]{bhowmik19}, we consider this feature, which could indicate the ring ansae, as marginally significant at best. The surface brightness profile from the ADI image has a similar shape overall but is $\approx$40\% lower than in the other images. Overall, this is consistent with the results of our injection-recovery tests and the match between the other three methods for the \hd\ dataset provides further confidence in the reliability of the surface brightness profiles derived here.

Both the total and polarized intensity profiles are highly symmetrical about the star, with differences never exceeding 20\% at any stellocentric distance. This is in contrast with past claims of significant asymmetries in the inner 1\arcsec\ \citep[e.g.,][]{schneider2005discovery,currie2012keck}. Subsequent analyses suggested that PSF subtraction artefacts could be misinterpreted as physical asymmetries \cite{esposito2013modeling}. In agreement with \cite{bhowmik19}, we do not recover the local ``gaps" observed at $\approx$0\farcs7 in total intensity by \cite{asensio2016polarimetry}. Instead, the polarized intensity profile plateaus at the location of these putative gaps and we conclude that the PSF subtraction method employed by these authors amplified these features into apparent surface brightness deficits.

All profiles share a steep decline outside of $\approx$1\arcsec, i.e., in the disk halo. We performed power law fits and found that the surface brightness profile follows approximately $r^{-4}$ and $r^{-5}$ in polarized and total intensity, respectively. These are in reasonable agreement with previous studies \citep{boccaletti2012morphology,currie2012keck,esposito2013modeling}, although the limited field-of-view of our observations significantly reduces the precision of our estimates. Inside of a marked inflection point around the disk ansae, the total intensity brightness profile follows $r^{-1.5}$, with suggestive evidence for a gradual steepening towards the smallest projected separations. Again, this is in reasonable agreement with past studies of the system.

Contrary to the total intensity surface brightness profile, the polarized intensity profile displays a broad plateau over the 0\farcs4--0\farcs9 range. The outer edge of this plateau lies $\approx15$--20\,au outside the ring radius inferred in \S\,\ref{subsec:morph}. This may indicate that the ring has a non-negligible radial extent, an issue that we will revisit in \S\,\ref{sec:model}. Inside of this plateau, the polarized surface brightness profile follows $r^{-1.5}$, similar to the total intensity profile. While it could be tempting to interpret the break at 0\farcs4 as an indication for a secondary ring (with a radius of $\approx$50\,au), the absence of any "kink" in the disk spine at that location argues against this scenario. Instead, the central peak in polarized surface brightness must be due instead to sufficiently strong forward scattering to overwhelm the polarization decline inherent to the smallest scattering angles.

\begin{figure}
\epsscale{1.17}
\plotone{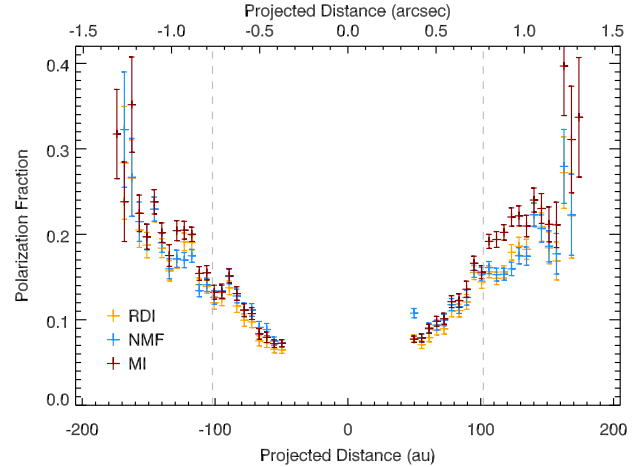}
\caption{$H$-band polarization fraction across the \hd\ disk as a function of stellocentric distance. Three of the PSF subtraction methods are used to estimate systematic uncertainties associated with this process. The ADI subtraction is not considered here since it systematically under-evaluates the disk surface brightness. The vertical dashed lines indicate the disk radius as derived from the geometric fit to the disk spine. \label{fig:polar_dist}}
\end{figure}

Combining the total and polarized intensity surface brightness profile, we compute the polarization fraction along the disk spine. The results are shown in Figure\,\ref{fig:polar_dist}. We observe a steady rise, from about 7\% at a projected distance of 0\farcs35 from the star, to 15\% at the ring ansae, and up to 20--30\% at 1\farcs3. Our results match well with those obtained by \cite{asensio2016polarimetry}. This degree of linear polarization is within the range of near-infrared observations of debris disks \citep{tamura06, perrin15, draper16, esposito18}.

\begin{figure*}
\epsscale{1.17}
\plottwo{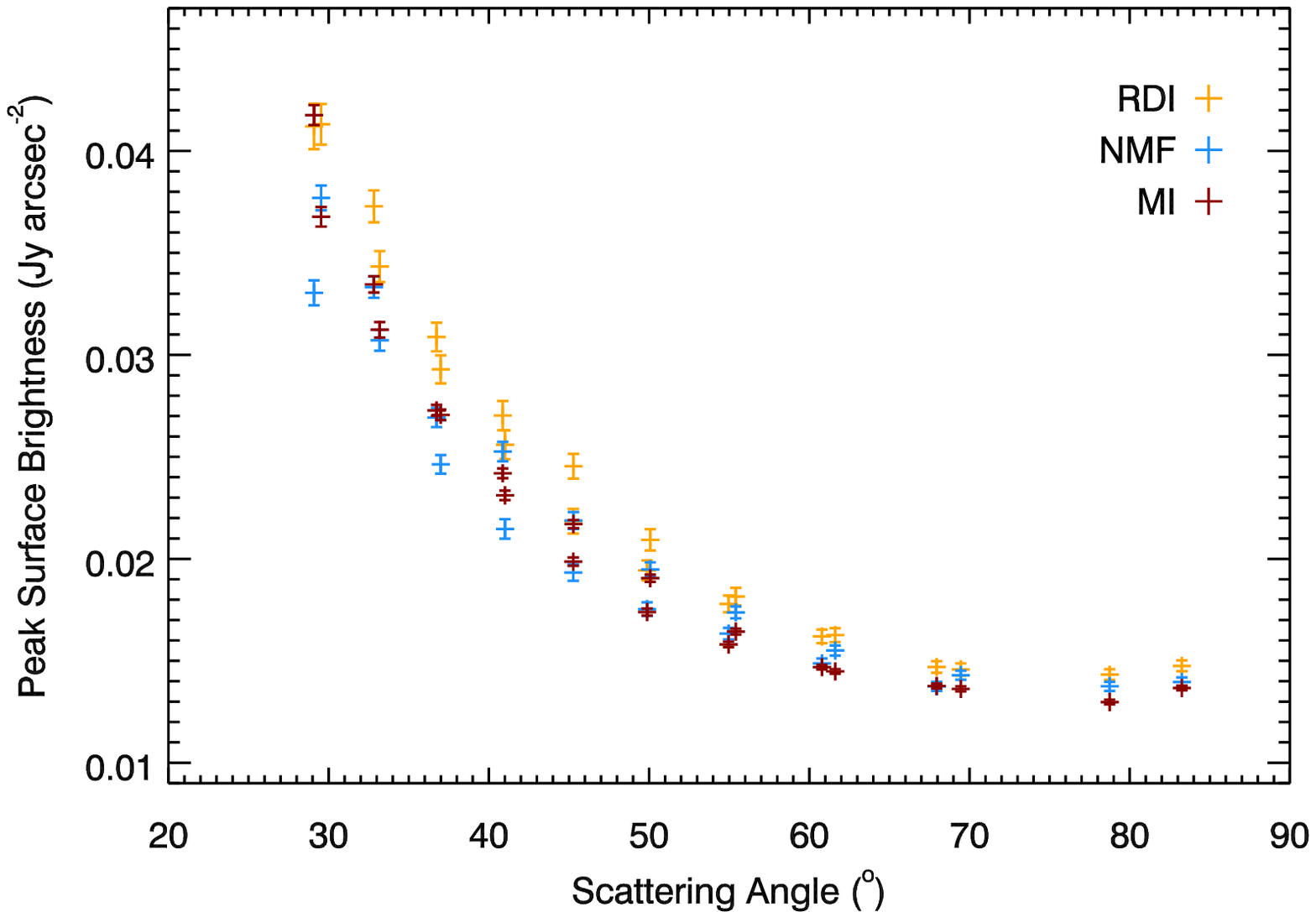}{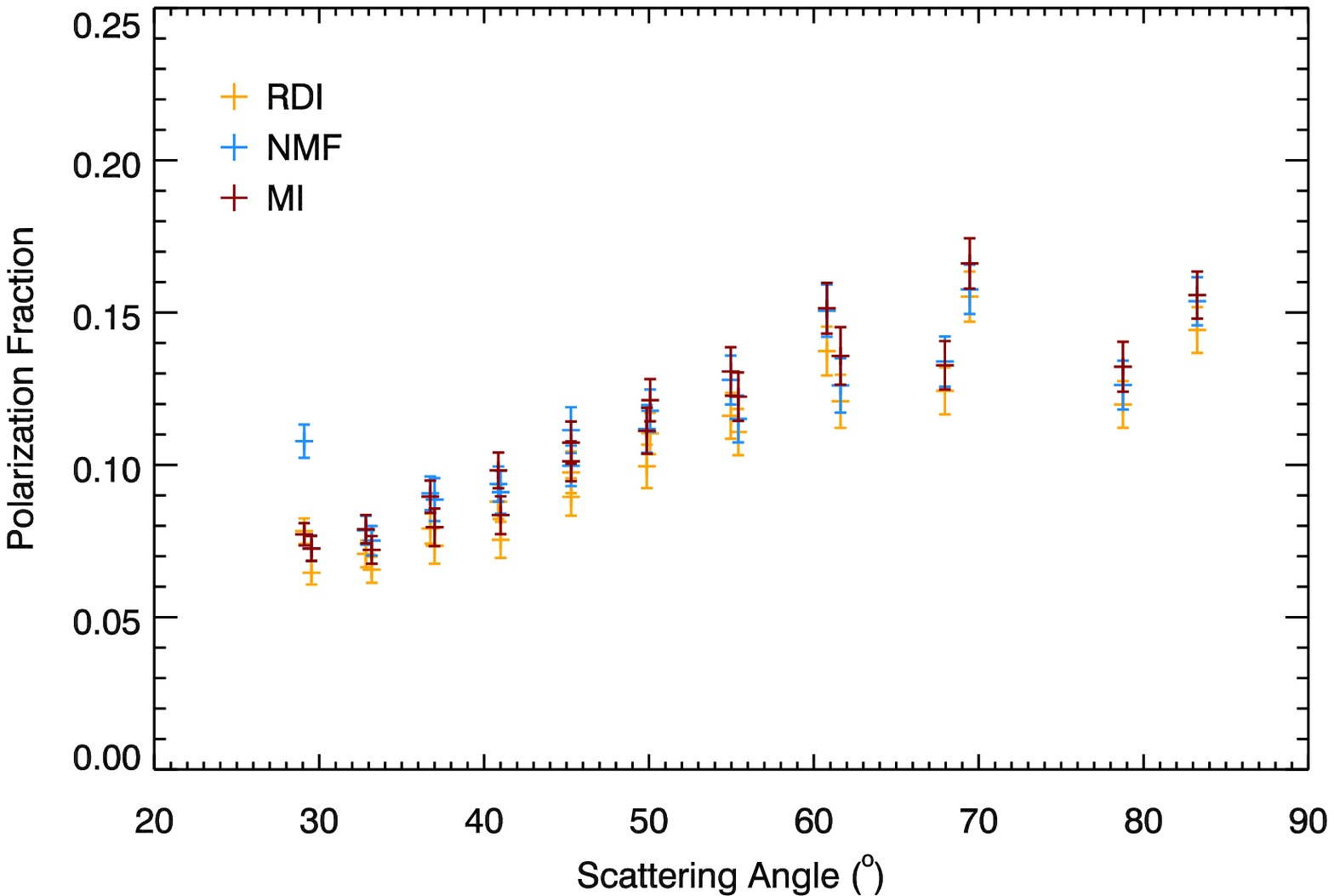}
\caption{$H$-band SPF (left) and polarisability curve (right) of the \hd\ disk, using the best estimate of the ring geometry. The color symbols indicate the different PSF subtraction methods. 
The ADI PSF subtraction is affected by a significant, and likely position-dependent, self-subtraction which precludes estimating the underlying surface brightness profile without a dedicated forward modeling approach.
The two sides of the disk are plotted separately. However, the fact that the best-fit offset of the ring center is small results in nearly identical scattering angles being estimated on each side of the star, except close to the ansae. \label{fig:polar_scattang}}
\end{figure*}

To constrain the properties of the dust grains in the \hd\ disk, we need to extract the scattering phase function (SPF) and the polarisability curves, i.e., the dependency of the total intensity and degree of linear polarization as a function of scattering angle. Under the assumption of a narrow ring, there is a simple analytical transformation between the projected position of a point along the ring spine into a scattering angle. We therefore use the best fit geometry derived above from the polarized intensity image to estimate the scattering angle for every point along the spine out to the location of the ring ansae. The resulting curves are shown in Fig.\,\ref{fig:polar_scattang}. One caveat in this process is that close to the ansae, the backside of the disk can contribute significantly to the observed surface brightness since the difference in scattering angle between the front and back side is small, leading to limb brightening. Therefore, we expect that the true SPF of \hd\ declines more steeply at the largest scattering angles than we measure here. On the other hand, if the polarisability curve is symmetric about 90\degr\ \citep[as seen in cometary dust, e.g.,][]{frattin19}, this effect would cancel out when we compute the polarization fraction and we thus expect the polarisability curve we derive to be more robust.

The $H$-band SPF we derive for \hd, which declines by a factor of about 2.5 between scattering angles 30\degr\ and 60\degr, where contribution from the back side should be minimal based on the disk's curved spine, is consistent with the nearly-universal SPF observed for Solar System, debris disks and protoplanetary disks dust populations \citep{hughes2018debris}. On the other hand, it clearly deviates from that observed in the HR\,4796 debris disk \citep{perrin15, milli17} as the latter shows a minimum at a scattering angle of $\approx60$\degr. There are too few polarisability curves published to date for debris disks to draw definitive conclusion, but the curve we obtain for \hd\ is much more consistent with that observed in the HD\,35841 system \citep{esposito18} than in HR\,4796 \citep{perrin15}.

\section{Modeling} \label{sec:model}

We now proceed to evaluate the physical properties of the dust grains: in particular, the grain size distribution and composition (\S\ref{subsec:dustprops}). We then perform a consistency test of our initial narrow ring assumption by directly fitting the disk images based on the derived dust properties (\S\ref{subsec:image_modeling}). In principle, a simultaneous fit to the GPI images, with all dust properties and disk geometry parameters left free to vary, represents the most direct approach. However, in cases where models suffer from systematic shortcomings, this can lead to a false sense of success, whereas the multiple-step approach used here allows us to disentangle which assumptions are not verified in our analysis. The general implications of our modeling results are discussed in \S\,\ref{sec:discus}.

The NMF, RDI and MI PSF subtraction methods yield consistent surface brightness profiles and thus SPF and polarisability curves. We select the MI-based results for this analysis since it offers the smallest inner working angle. Furthermore, this method intrinsically yields much smaller systematic residuals (see Figure\,\ref{fig:psfsub}), suggesting that the pixel-to-pixel uncertainties are more likely to be mostly random in nature. 

\subsection{Dust Properties Analysis} \label{subsec:dustprops}

\subsubsection{Modeling setup}

Here we wish to reproduce the SPF and polarisability curves derived from our observations of the \hd\ disk. We adopt the Mie model, valid for compact, spherical dust grains of homogeneous composition. We note that observations of both laboratory and astrophysical dust populations suggest that this assumption is not optimal \citep[e.g.,][]{pollack80, hedman15, milli17}. However, it is computationally tractable in the context of large dust grains, a problem not yet solved for grain aggregates that are likely to represent a better model of astrophysical dust \citep[e.g.,][]{arnold19}. 

Two components are necessary to build a dust model: the grain size distribution and the dust composition. We follow standard approaches and assume a power law size distribution, $N(a)\,\mathrm{d}a \propto a^{-\eta}\,\mathrm{d}a$ ranging from $a_{\mathrm{min}}$ to $a_{\mathrm{max}}$. Collisional cascade models predict a size distribution with $\eta \approx 3.5$ \citep[e.g.,][]{dohnanyi1969collisional, marshall17}, although deviations from a pure power law are likely \cite[e.g.,][]{thebault07}. On the other hand, the dust composition is a more challenging issue to handle. It is most often addressed either as a fixed, presupposed, composition or as a mixture with variable proportions of several individual compositions (using effective medium theory). While easiest to implement, the first approach can lead to significantly biased results or, worse, a lack of model that fits the data well if an incorrect composition is picked. The dust mixture suffers from increasing the number of free parameters and, in the worst case scenario, a critical component may be left unexplored. For instance, \cite{rodigas15} consider 19 different dust compositions, plus vacuum to represent porosity, when modeling the HR\,4796 debris disk. Even then, only a subset of the data is well fit by the resulting model. To circumvent these issues, we adopt a more direct approach, which consists of fitting for the material's complex refractive index $m = n + i\,k$, as this is the quantity from which Mie theory predicts the SPF and polarisability curve. A similar approach was adopted by \cite{graham2007signature} in their modeling of the polarised scattered light imaging of the AU\,Mic debris disk and was instrumental in identifying the need for a large dust porosity in that system.

Because \hd\ is nearly, but not quite, edge-on, we expect that the back side of the disk contributes to the signal close the ansae. To account for this effect in our models, and taking advantage of the absence of a lateral offset of the central star, we modify the Mie-computed SPF by adding the contributions of the front and back sides using supplementary scattering angles. Similarly, we compute the average of the front and back side polarized intensity signals to obtain the final version of the model polarisability curve. Approximating the disk has being exactly edge-on, we perform this correction at all scattering angles, noting that the correction is only significant close to the ansae. In addition, because monochromatic calculations can experience interference fringes in model SPF and polarisability curves, we compute the Mie models at 9 wavelengths spanning the bandpass of the GPI $H$ band filter and average the resulting curves over the wavelength prior to computing the model likelihood. Finally, we normalize all SPFs to their average value in the 40--60\degr\ range of scattering angle in order to focus on the shape of these curves.

We set up three independent parallel-tempered MCMC chains using the {\tt emcee} package \citep{foremanmackey2013emcee}. The first one fits only the \hd\ SPF, the second only the polarisability curve, and the third fits both curves simultaneously. In all cases, the model likelihood is based on a standard $\chi^2$ test between the observed and modeled curve. Each of these runs includes 2 temperatures and 50 walkers. Walkers are initially distributed using uniform priors spanning the ranges indicated in Table\,\ref{tab:dust_fit}. We remove the first 80\% of each chain as a burn-in and use the final 20\% to obtain values reported in Table\,\ref{tab:dust_fit}, with 1440 iterations kept after burn-in for our SPF fit, 3284 iterations for our polarisability fit, and 3784 iterations for our joint fit. Inspection of the movements of walkers in the parameter space confirm that the chains are well converged.

\subsubsection{Results}

\begin{table*}
\caption{Best-fitting dust properties based on the SPF alone, the polarisability curve alone, and both curves simultaneously. The range explored for each quantity in indicated in the second column. Upper and lower limits are reported at the 95\% confidence level.\label{tab:dust_fit}}
\begin{center}
\begin{tabular}{|c|c|ccc|cccc|}
\hline
Parameter & Prior & \multicolumn{3}{c|}{Best-fitting Model} & \multicolumn{4}{c|}{Median $\pm 1\sigma$} \\
 & Range & SPF & Polar. & Joint & SPF & Polar. &  \multicolumn{2}{c|}{Joint}\\
 & & & & & & & Peak\,1 & Peak\,2 \\
\hline
$\log (a_{\mathrm{min}}\, [\mu\mathrm{m}])$ & [-1 .. 1] & -0.08 & -0.10 & -0.14 & -0.09$^{+0.01}_{-0.01}$ & -0.11$^{+0.01}_{-0.58}$ & -0.143$^{+0.004}_{-0.004}$ & -0.564$^{+0.002}_{-0.002}$ \\
$\log (a_{\mathrm{max}}\, [\mu\mathrm{m}])$ & [1 .. 3] & 2.84 & 0.341 & 2.99 & 2.01$^{+0.68}_{-0.67}$ & 2.07$^{+0.61}_{-0.66}$ & 
$\geq2.50$ & $\leq1.05$ \\
$\eta$ & [2 .. 5] & 4.14 & 3.56 & 3.52 & 4.21$^{+0.13}_{-0.13}$ & 3.54$^{+0.11}_{-0.20}$ & 3.516$^{+0.008}_{-0.01}$ & 
3.52$^{+0.02}_{-0.02}$ \\
$n$ & [1 .. 5] & 3.31 & 2.64 & 3.78 & 3.30$^{+1.07}_{-0.10}$ & 4.13$^{+0.67}_{-1.51}$ & 3.78$^{+0.03}_{-0.03}$ & 
3.49$^{+0.06}_{-0.06}$ \\
$\log k$ & [-7 .. 1] & -6.16 & -1.29 & -1.44 & 
$\leq -2.83$ & 
-1.37$^{+0.11}_{-0.09}$ & -1.44$^{+0.01}_{-0.01}$ & 
-0.77$^{+0.02}_{-0.02}$ \\
\hline
\end{tabular}
\end{center}
\end{table*}

\begin{figure*}
\epsscale{1.17}
\plottwo{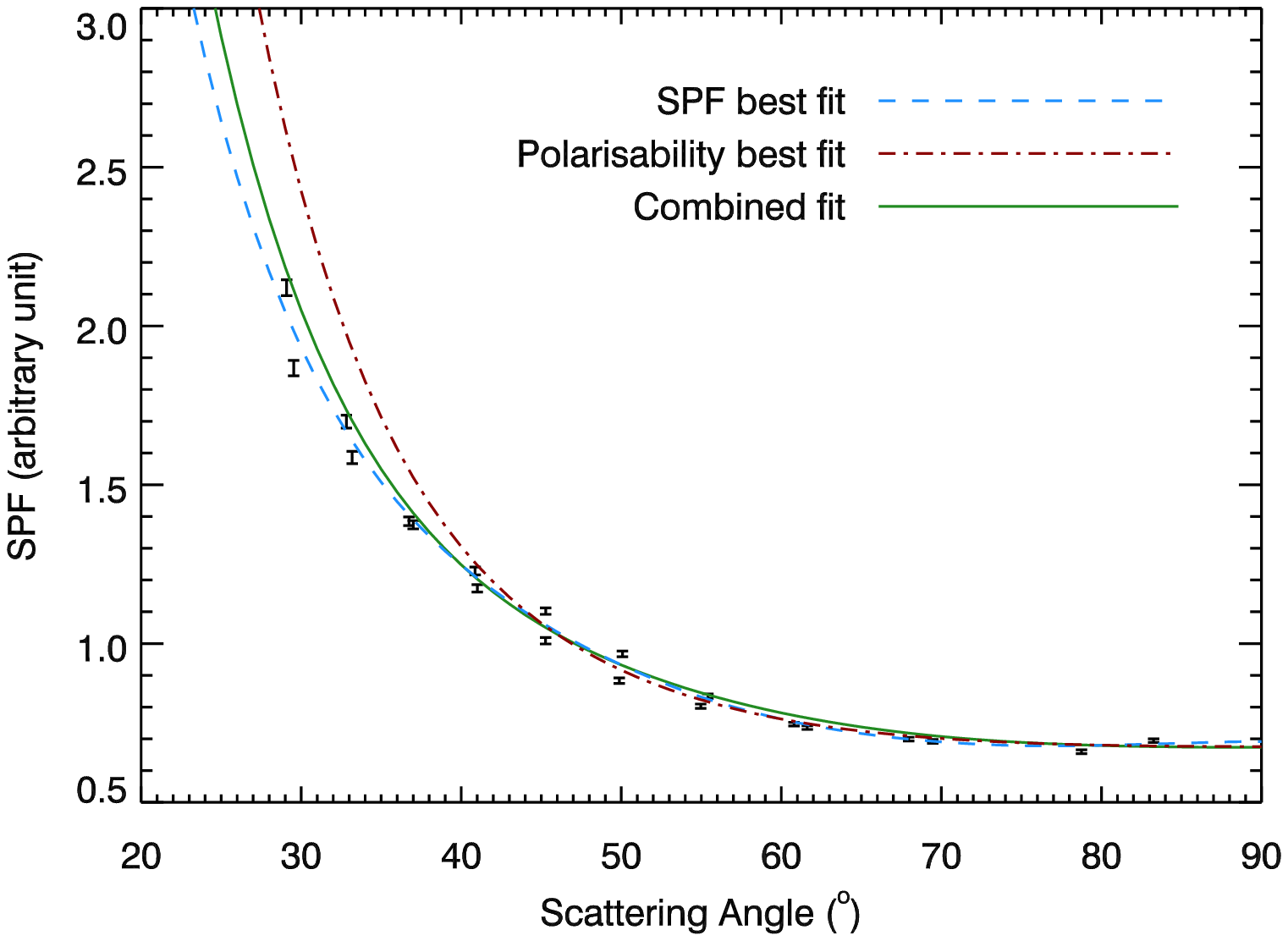}{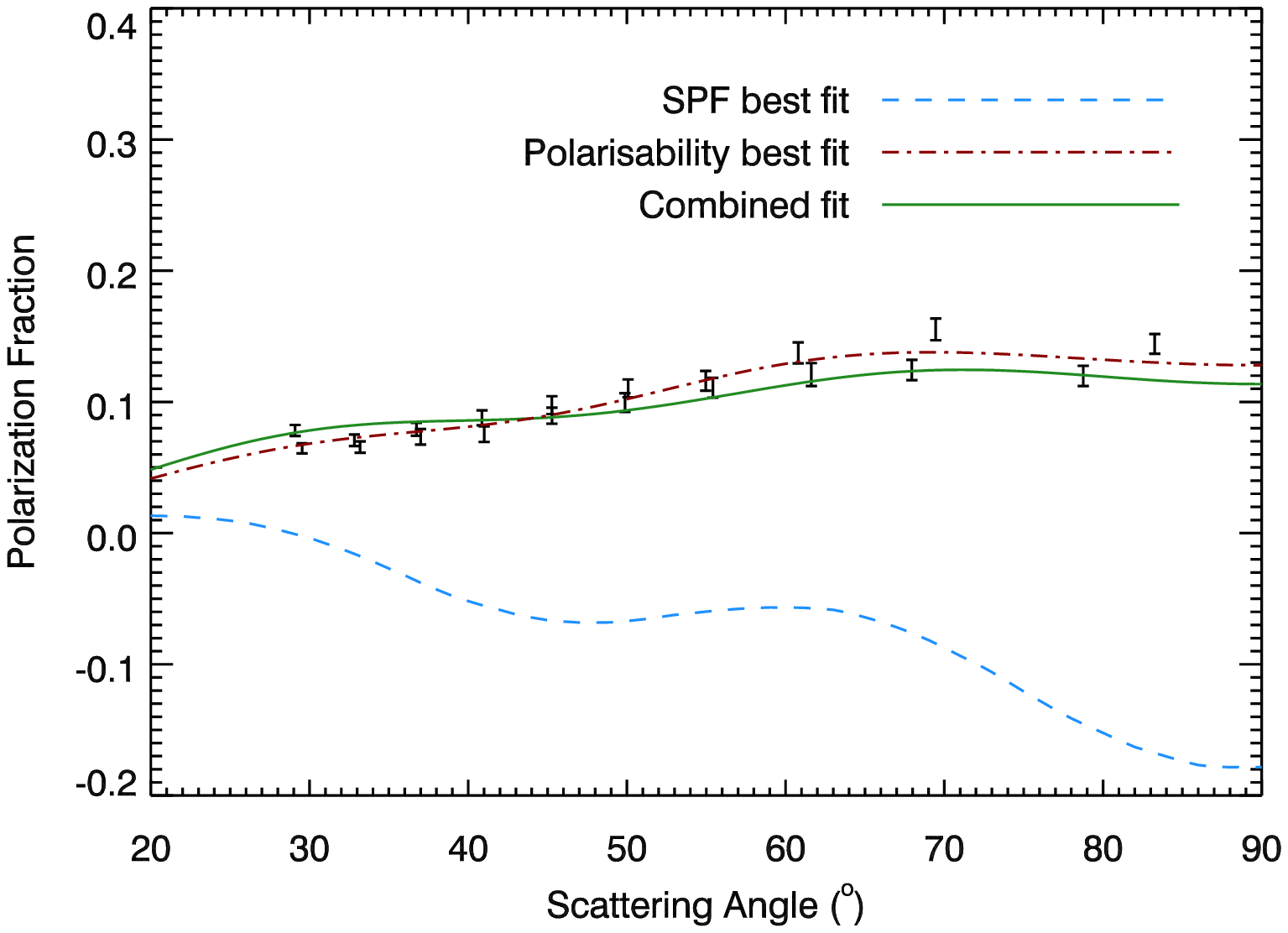}
\caption{Observed and modeled SPF (left) and polarisability (right) curves. The model curves are modified to account for the superimposition of the front and back sides of the disk. Observed quantities, as derived from the MI total intensity image, are shown as black errorbars while the colored curves represent the best fit to the SPF (red dot-dashed), to the polarisability curve (blue dashed) and to both curves simultaneously (solid green).\label{fig:polfrac_scattint}}
\end{figure*}

Our final best-fitting model from each of these runs is displayed in Figure\,\ref{fig:polfrac_scattint}, with parameters described in Table\,\ref{tab:dust_fit}. While both the observed SPF and polarisability curves are reasonably well reproduced when either quantity is fit separately, the corresponding reduced $\chi^2$ values are 11.8 and 2.3, respectively. These imperfections are driven by the fact that the SPF (and to a lesser degree the polarizability curve) measured on the NE and SW sides of the disk are formally inconsistent with one another, and the best fit model is a compromise between both sides. As a result, the formal parameter uncertainties derived from the MCMC process are likely underestimated. Nonetheless, Figure\,\ref{fig:polfrac_scattint} illustrates that our best-fitting models reproduce the overall shape of both the SPF and polarisability curves, suggesting that the values of the best-fitting parameters can be considered as reliable.

Although the model parameters for all three fits (``SPF only", ``polarisability only", ``combined") are significantly different, all three model SPFs are similar to the observed one (left panel in Figure\,\ref{fig:polfrac_scattint}). This suggests that the SPF of the \hd\ dust disk is consistent with a large swath of the parameter space, indicating that this quantity has limited discriminatory power as far as dust properties are concerned. This is qualitatively consistent with the observations that many astrophysical dust population share similar scattering SPFs \citep{hughes2018debris}. On the other hand, the polarisability curve may be significantly more constraining, since the ``SPF only" dust model is highly inconsistent with the observed polarisability curve. Specifically, due to its much steeper size distribution and very small value of the imaginary part of the refractive index, that model predicts a negative polarization at most relevant scattering angles, i.e., polarisation vectors that are radially organized instead of ortho-radial. This is readily excluded by the fact that the Stokes $Q_\phi$ map shows only positive signal along the disk. Unsurprisingly, the combined fit resembles the ``polarisability only" fit much more than the ``SPF only" fit.

\begin{figure*}
\plotone{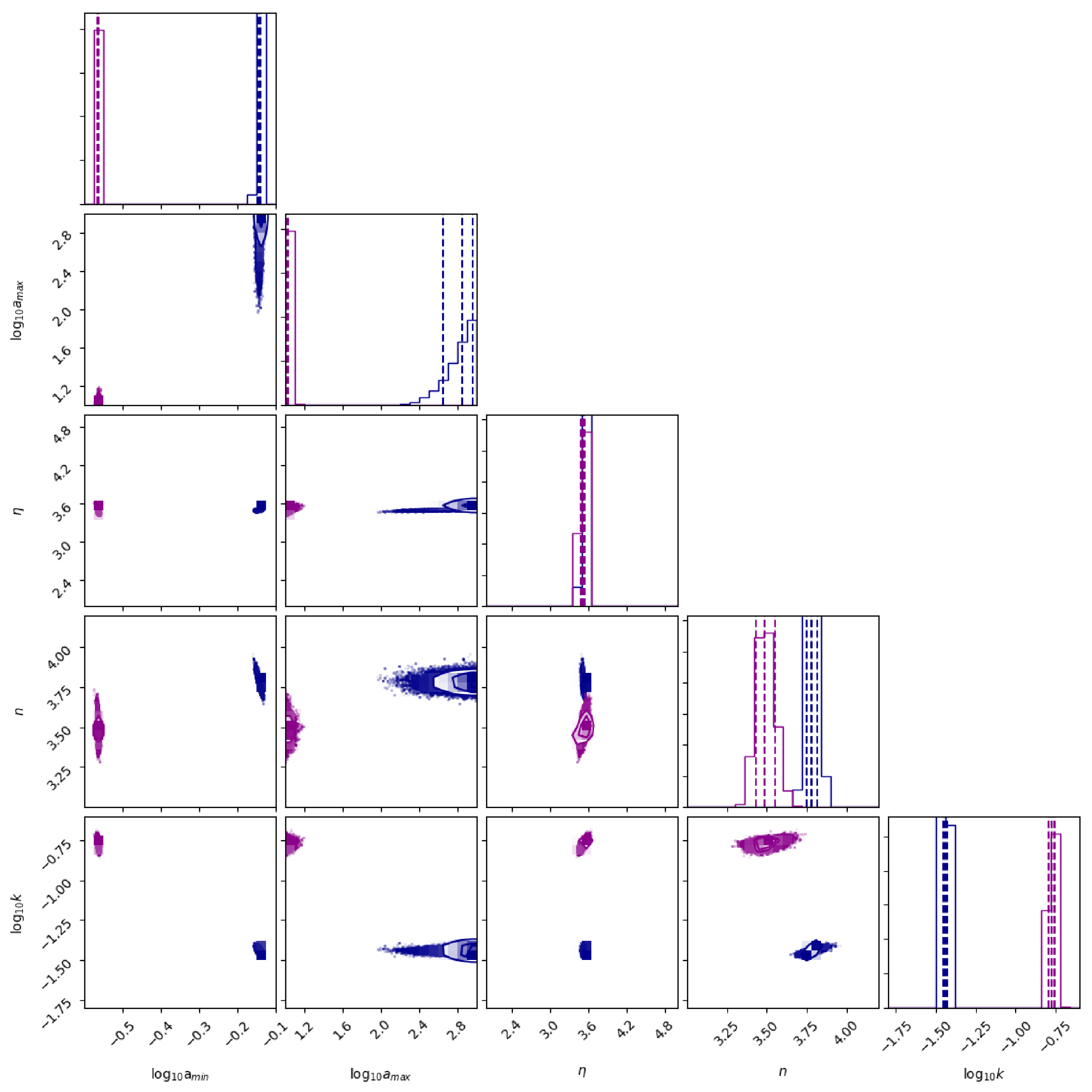}
\caption{Posterior distributions for the joint SPF and polarisability dust properties fit. The two distinct families of acceptable models (shown in blue and purple, respectively) are considered separately in extracting the confidence intervals presented in Table\,\ref{tab:dust_fit} (as Peak\,1 and 2, respectively). Vertical dashed lines marke the 16-, 50- and 84-percentile values for each parameter and family of models.}
\label{fig:dust_peak}
\end{figure*}

Turning our attention to the best fitting model parameters, we first note that the "combined" fit lead to two distinct families of models, as illustrated in Figure\,\ref{fig:dust_peak}. The family characterized by a large value of $a_\mathrm{max}$, which is referred to as ``Peak\,1" in Table\,\ref{tab:dust_fit}, is consistent with both the ``SPF only" and ``polarisability" fits and we thus consider it as the most plausible model. Besides this consistency, the other family of models is characterized by a very narrow grain size distribution (in particular, $a_\mathrm{max} \lesssim 11\,\mu$m at the 95\% confidence level) that seems physically unlikely. In the remainder of the analysis, we focus on the first family of models.

All three fits yield consistent minimum grain sizes, $a_\mathrm{min} \approx0.8\,\mu$m. Conversely, we find that the maximum grain size is constrained to be large, with a 95\% confidence level lower limit of $440\,\mu$m. Finally, we note that, while the ``SPF only" and ``polarisability only" fits each constrain the size distribution power law index well, they yield inconsistent values: $\eta \approx 4.2$ and 3.5, respectively. The ``combined" fit favors the latter value, which is consistent with collisional models.

In both the ``SPF only" and ``polarisability only" fits, we find multimodal posteriors spanning a large fraction of the explored range for the real part of the refractive index but with little overlap between one another, indicating ambiguities in the fit. Striving to achieve a compromise between the two observed quantities, the ``combined" fit has a significantly narrower posterior, $3.5 \lesssim n \lesssim 3.8$. The imaginary part of the refractive index also reveals significant tension between the ``SPF only" and ``polarisability only" fits: the former yields an upper limit on $k$, $\log k \lesssim -2.8$, while the latter has well constrained posterior, $\log k \approx -1.4\pm0.1$. The ``combined fit" posterior prefers the latter solution with a secondary peak at $\log k \approx -0.8$. 

We defer the interpretation of the results of our dust fitting to \S\,\ref{sec:discus}. For now, we note that, while the SPF and polarisability fit leave some unsolved ambiguities and tensions, the best-fitting model yields an acceptable fit to both quantities. In turn, this allows us to fix the dust properties and perform image fitting to assess the geometrical properties of the disk. 

\subsection{Image modeling} \label{subsec:image_modeling}

\subsubsection{Modeling setup}

To model the GPI total intensity and Stokes $Q_{\phi}$ images, we use the best-fitting combined dust model derived in the previous section and explore the geometrical structure of the disk. We use the {\tt MCFOST} radiative transfer code \citep{pinte2006monte} to produce synthetic scattered light images. We model the debris disk density structure with the widely used functional form 
$$
\rho(r,z) \propto \frac{e^{-(\frac{\left|z\right|}{h(r)})^{\gamma_\mathrm{vert}}}}
{\sqrt{\big(\frac{r}{r_\mathrm{c}}\big)^{-2\gamma_\mathrm{1}} + \big(\frac{r}{r_\mathrm{c}}\big)^{-2\gamma_\mathrm{2}}}},
$$
following \cite{augereau1999hr}. The critical radius, $r_\mathrm{c}$, marks the transition between two power law density regimes (with indices $\gamma_\mathrm{1} > 0$ and $\gamma_\mathrm{2}< 0$, respectively). We set $\gamma_\mathrm{vert}=2$ to yield a Gaussian vertical profile and a bow-tie shape for the disk, i.e., a constant $h/r$ ratio. We further restrain the radial extent of the disk with inner and outer hard edges at radii $r_\mathrm{in}$ and $r_\mathrm{out}$, mostly for computational purposes.

Given a disk geometry and a set of dust properties, {\tt MCFOST} produces a full Stokes synthetic datacube with pixel scale, orientation and field of view set to match our GPI observations. The Stokes $Q$ and $U$ maps are converted to a Stokes $Q_\phi$ image, while the star is masked out of the Stokes $I$ image, before both are convolved by the instrumental PSF as estimated by the satellite spots. We then mask regions that are closer than the inner working angle of the Stokes $I$ image to only consider the same pixels that were used in deriving the SPF and polarisability curve in the previous section. We also set an outer radius of 1\farcs6, outside of which no trustworthy data are available. Finally, we mask out pixels that lie more than 0\farcs35 from the disk spine to ensure that the fitted region include both disk-dominated and background-dominated pixels. A likelihood is then computed using a pixel-by-pixel $\chi^2$ calculation. Exploration of the parameter space is conducted through three independent MCMC processes -- one fit for Stokes $I$, a second for Stokes $Q_\phi$, and a third combined fit -- each using 3 temperatures and 100 walkers. Our final results again include only the final 40-45\% of each MCMC chain, when the chains had visually achieved convergence (in total, this includes 1660, 2420 and 1380 iterations for our Stokes $I$ fit, Stokes $Q$ fit, and joint fit, respectively). Consistent with dust modeling conducted above, we adopt the MI PSF-subtracted total intensity image of the disk.

In this aspect of our modeling, the geometrical free parameters are the disk inclination ($i$), the critical radius ($r_\mathrm{c}$), the volume density power law indices ($\gamma_\mathrm{1}$, $\gamma_\mathrm{2}$), the reference scale height ($h_0$, defined at $r_0 = 100\,\mathrm{au}$), and the disk inner radius ($r_\mathrm{in}$). Given the large halo that extends well beyond the GPI field-of-view, we cannot constrain the disk outer radius with our data and thus set $r_\mathrm{out} = 200$\,au. Finally, we set the total disk mass ($M_\mathrm{d}$) as a free parameter that defines the total amount of dust in the system, based on a representative grain density of 3.5\,g.cm$^{-3}$. So long as the disk remains optically thin, this acts as a simple multiplicative factor that serves to adjust the absolute surface brightness of the model to the observed one. We initialize $\gamma_\mathrm{1}$ and $\gamma_\mathrm{2}$ with uniform distributions, and all other free parameters are assigned a Gaussian prior, either based on our empirical geometrical analysis ($i$, $r_\mathrm{c}$, $h_0$, from \S\,\ref{subsec:morph}) or assuming a conservatively broad range ($r_\mathrm{in}$, $M_\mathrm{d}$). The explored ranges for each parameter are indicated in Table\,\ref{tab:geom_fit}.

\begin{table*}
\caption{Best-fitting geometrical properties based on the Stokes $I$ image, the Stokes $Q_\phi$ image, and both images simultaneously. The range explored for each quantity in indicated in the second column. The range explored for each quantity in indicated in the second column. Upper and lower limits are reported at the 95\% confidence level.\label{tab:geom_fit}}
\begin{center}
\begin{tabular}{|c|cc|ccc|ccc|}
\hline
Parameter & \multicolumn{2}{c|}{Prior Range} & \multicolumn{3}{c|}{Best-fitting Model} & \multicolumn{3}{c|}{Median $\pm 1\sigma$} \\
 & Initial & Full & $I$ & $Q_\phi$ & Joint & $I$ & $Q_\phi$ & Joint \\
\hline
$i$ [\degr] & 87$\pm$1 & [70 .. 90] & 88.84 & 88.65 & 88.74 & $88.88^{+0.01}_{-0.02}$ & $88.59^{+0.04}_{-0.05}$ & $88.75\pm0.02$ \\
$h_0$ [au] & 5$\pm$2 & [0.1 .. 10] & 0.10 & 0.10 & 0.10 & $0.13\pm0.02$ & $0.102^{+0.002}_{-0.001}$ & $0.11^{+0.04}_{-0.01}$ \\
$r_\mathrm{c}$ [au] & 100$\pm$20 & [50 .. 150] & 99.79 & 95.81 & 98.35 & $97.74^{+0.96}_{-0.79}$ & $93.87^{+1.21}_{-0.90}$ & $98.20^{+0.37}_{-0.56}$ \\
$r_\mathrm{in}$ [au] & 50$\pm$20 & [1 .. 100] & 50.50 & 7.38 & 39.85 & $51.8^{+2.2}_{-2.8}$ & $22.2^{+7.8}_{-3.0}$ & $41.8^{+2.6}_{-3.4}$ \\
$\gamma_\mathrm{1}$ & [0 .. 5] & [0 .. 5] & 4.01 & 3.08 & 3.42 & $3.77^{+0.23}_{-0.20}$ & $3.44^{+0.32}_{-0.31}$ & $3.37^{+0.11}_{-0.12}$ \\
$\gamma_\mathrm{2}$ & [-5 .. 0] & [-5 .. 0] & -4.79 & -4.96 & -4.96 & $-4.60^{+0.10}_{-0.12}$ & $\leq-4.84$ & $\leq-4.87$ \\
$\log_{10} (M_\mathrm{d} [M_\odot])$ & -9$\pm$2 & [-12 .. -4] & -7.60 & -7.31 & -7.57 & $-7.61\pm0.01$ & $-7.37^{+0.01}_{-0.02}$ & $-7.60^{+0.01}_{-0.02}$ \\
\hline
\end{tabular}
\end{center}
\end{table*}

\subsubsection{Results}

\begin{figure*}
\plotone{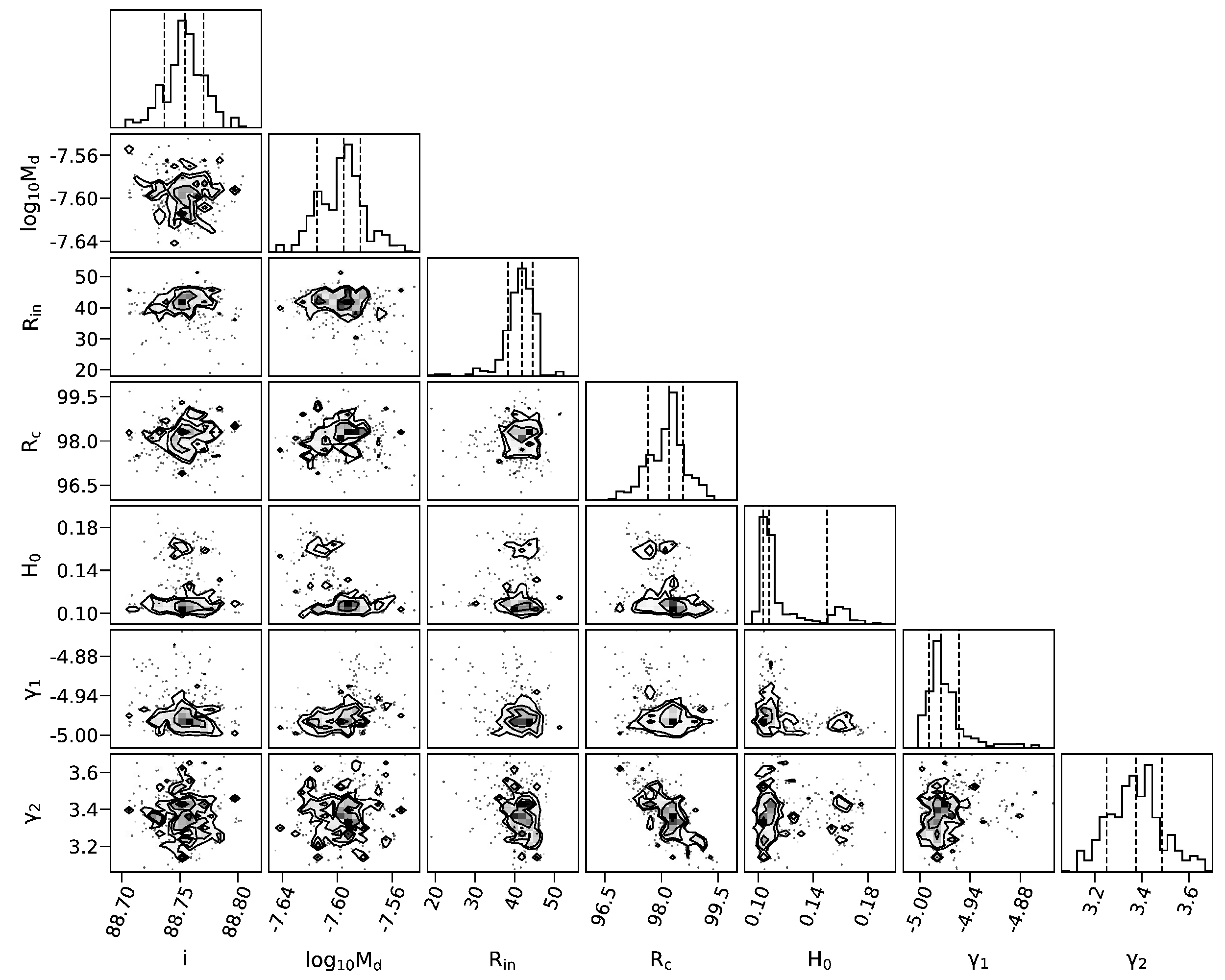}
\caption{Posterior probability distribution for all free parameters in the joint fit to Stokes $I$ and $Q_\phi$ intensity images.}
\label{fig:corner}
\end{figure*}

\begin{figure*}
\plotone{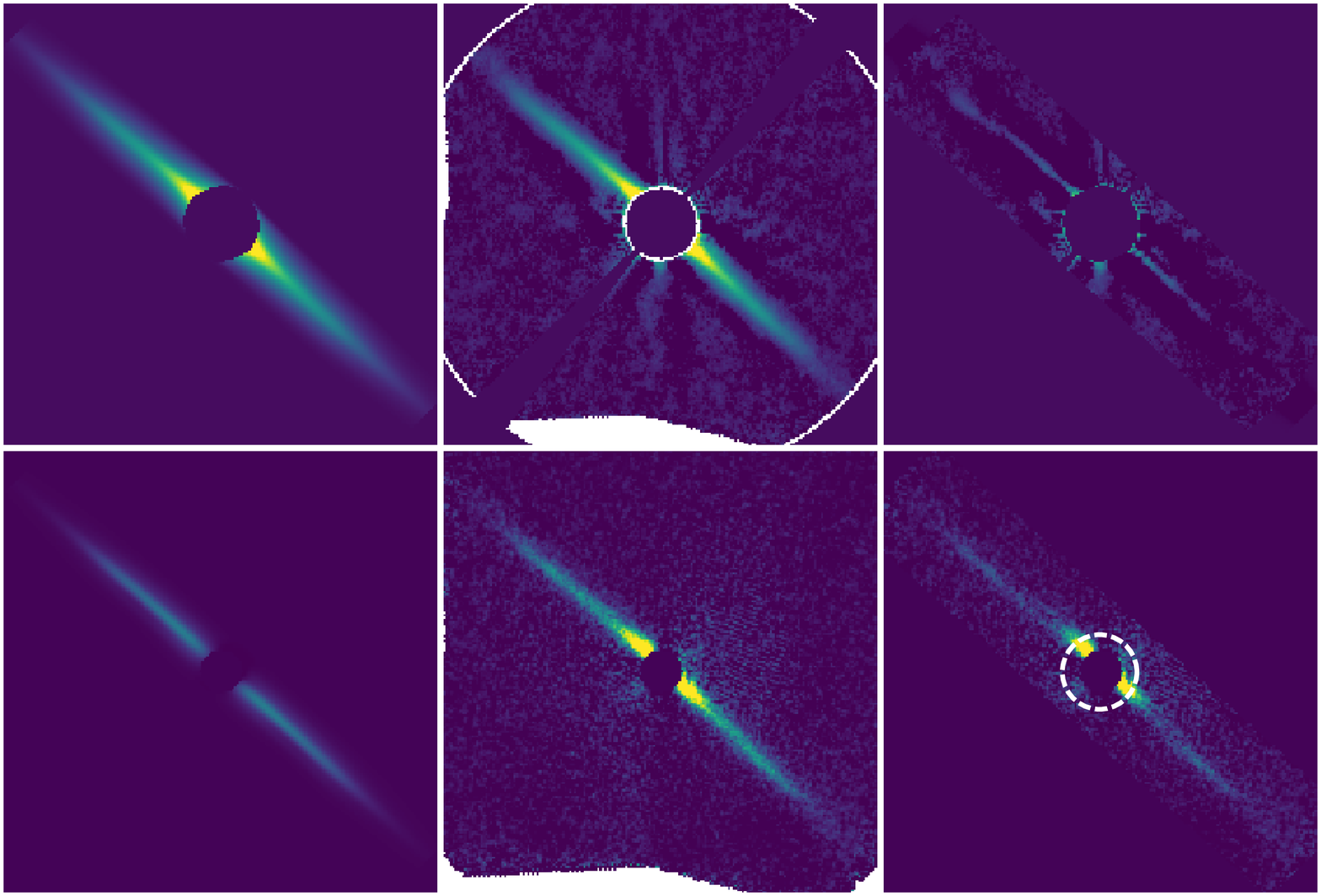}
\caption{Best-fitting total and polarized intensity images for {\hd} (top and bottom rows, respectively). From left to right in each row is the model image, $H$-band data image, and residuals all on the same scaling (square root stretch for Stokes $I$, linear stretch for Stokes $Q_\phi$). In the bottom right panel, the dashed circle indicate the inner working angle used in estimating the SPF and polarizability curve. Data inside of that circle are not included in the dust property fit and, consequently, are not included in the image fit either to prevent any bias. The residuals are shown here for visual display only.}\label{fig:bestfit_images}
\end{figure*}

The results of our MCMC chain are summarized in Table \ref{tab:geom_fit}. Figure \ref{fig:corner} displays the full posterior distribution for all parameters in the combined fit and Figure \ref{fig:bestfit_images} shows the model images for the overall best-fitting model. While the posteriors appear multi-modal, particularly for $H_0$, we inspected the movement of the walkers in the MCMC chains to confirm that the chains had been decoupled from their initial state. The results of fitting separately the Stokes $I$ and $Q_\phi$ images are mostly similar to those of the combined fit, although the scatter in some of the parameter values exceed the nominal uncertainties from the MCMC chains. For instance, fitting the polarized intensity image yields a 10\% smaller disk radius and an 0\fdg2 lower inclination. In the following we consider all three separate fits holistically in our analysis.

Overall, the best-fitting model reproduces well the observed images, at least within the region used in the likelihood function. However, given the high signal-to-noise of our dataset, the residuals are statistically significant, indicating that the model has some shortcomings. In particular, both the Stokes $I$ and $Q_\phi$ residual maps reveal a thin trace along the disk spine, suggesting that the model is slightly too extended vertically. Since we have allowed the vertical thickness of the disk to be very small ($h/r$ as low as 0.1\%), it is possible that this is due to our use of a slightly too broad instrumental PSF. We also find systematic residuals in the Stokes $Q_\phi$ image at the location of the ring ansae. This may be a consequence of imperfection in the dust scattering properties as derived in the previous section. Furthermore, we also note that the best fitting model underpredicts the polarized intensity in the immediate vicinity of the inner working angle of that image, a region not included in the fit. Finally, there are marginally significant positive residuals in the total intensity image outside of the ring radius. This is likely due to the lack of treatment of the halo in our model, although we stress that only the region within the ring radius is strongly detected in our data. Altogether, in spite of these limitations, we consider that the quality of the fit is sufficient to warrant a discussion of the main results from our exploration of the parameter space.

All geometric parameters are well constrained in the fit, except for $r_\mathrm{in}$ as a consequence of the steep inner surface density profile. Considering first the combined fit (to the Stokes $I$ and $Q_\phi$ images simultaneously), we derive an inclination of $i\approx$88\fdg7$\pm$0\fdg1, which is about 0\fdg5 higher than the best fit value obtained in the geometrical analysis in Section\,\ref{subsec:morph}. This is an indication that either our model is imperfect, or some of the assumptions that we used in deriving empirically the ring geometry are incorrect. This is further supported by the fact that we find a rather broad ring, with an inner radius 2--4 times smaller than $r_c\approx100\,\mathrm{au}$. Nonetheless, the latter is consistent with the ring radius we had derived in Section\,\ref{subsec:morph}. Although dust extends over a broad range of stellocentric distances, the power law volume density profiles are relatively steep ($\gamma_1 \approx 3.5$ and $\gamma_2 \lesssim -4.5$). The surface density profile is characterized by a FWHM of about 40\,au. This $\approx40$\% radial width is uncomfortably high to fully validate the narrow ring approximation of our initial geometric fitting and derivation of the SPF and polarizability curves. Nonetheless, the effect of this width is to blur the location of the ring spine and the dependencies on scattering angle, not to systematically bias these. We therefore expect our prior estimates to be representative of the true quantities, which is supported by the good match in the mean ring radius, for instance.

The surprisingly small disk scale height ($h/r \approx 0.2$\%) appears to contradict our finding that the disk is marginally resolved along the vertical direction (see Section\,\ref{subsec:morph}). Aside from the possibility that the PSF we used in the image modeling may not be a perfect match to the HD\,32297 dataset, this may be an indication that the vertical density distribution is not Gaussian. If the profile is more condensed in the center, e.g., following a Lorentzian or exponential profile, the assumption of a Gaussian profile in both our initial geometrical analysis and in radiative transfer modeling would overestimate slightly the vertical extent of the disk. Finally, the PSF subtraction process could have slightly attenuated the lower surface brightness regions away from the midplane in the total intensity image, thus leading to a similar effect. The fact that the fit to the polarized intensity image also favors a very small disk thickness rather points to other explanations, however.

Finally, the total dust mass, $M_\mathrm{d} \approx 0.01 M_\oplus$, is to be considered with caution as this quantity is strongly correlated with dust properties, particularly the minimum grain size and porosity. Since we have not attempted to fit for an actual composition, the true mean grain density is not a parameter of our model and is degenerate with the total mass. It is nonetheless interesting to note that this is much smaller than the dust mass derived from the millimeter emission of the system \citep[$\approx 0.6 M_\oplus$;][]{macgregor18}.

\section{Discussion} \label{sec:discus}

\subsection{System geometry}

The combination of high angular resolution and exquisite image fidelity enabled by GPI offers an opportunity to determine the ring geometry in a precise manner. This is further enhanced by the fact that the polarized intensity image does not require any PSF subtraction. It is thus interesting to note that the disk radius that we determined here, both from the direct geometric analysis and from the direct image fitting (Sections\,\ref{subsec:morph} and \ref{subsec:image_modeling}, respectively), is markedly smaller than has been found in past studies, around 100\,au compared to 130\,au. We emphasize that this difference is not a result of the updated distance to the system as we have already accounted for it. In other words, the angular radius of the ring we find is about 40\% smaller than previous studies. Notably, the two methods we employed rely on very different aspects of the data. Image fitting is inherently weighted by the signal-to-noise and, thus, by the local brightness of the disk, which places a different emphasis on different regions of the disk. The geometric approach, instead, is mostly independent of the surface brightness profile. Arguably, the latter is a more robust approach to determining the disk geometry. In particular, self- and over-subtraction effects have a much more direct influence on the surface brightness distribution and inadequately taking them into account is more likely to introduce biases than focusing on the spine of a nearly edge-on disk like \hd. The latter is now precisely traced as close as 0\farcs12 from the star \citep[this study; see also][]{bhowmik19} and the remaining dominant source of uncertainty may actually be the location of the star itself. In particular, \cite{bhowmik19}, who derived a disk radius from their SPHERE images that is consistent with past studies, suggest an offset of $\approx$0\farcs01 of the star in the direction perpendicular to the disk whereas our analysis reveals no such offset. While the nominal precision in the position of the star with instruments such as GPI and SPHERE is significantly better, this suggests that systematic uncertainties are not fully understood in these complex instruments.

Despite these systematic errors associated with scattered light images, the submillimeter emission of the system supports an 80--120\,au radial range for the ring \citep{macgregor18}. Even though the inferred surface density profile rises as roughly $r^2$ in their best fitting model, the $r^{-2}$ illumination dependency of impinging starlight yields a flat surface brightness profile and, thus, a roughly 100\,au radius for the scattered light ring. Similarly, the mid-infrared emission from the system suggests an inner disk radius of about 80--90\,au \citep{fitzgerald2007ring, moerchen200712}. While the scattered light images of the system may probe physically distinct grains, and there is evidence for mm-emitting dust in the halo surrounding the parent body belt \citep{macgregor18}, it seems implausible that the scatterers would be located exclusively outside of the parent body belt. This is definitely not the case in the well-studied, lower inclination, HR\,4796 system \citep{kennedy18}. We therefore conclude that the \hd\ ring is indeed centered at about 100\,au, as inferred from the modeling of our near-infrared image.

\begin{figure}
\epsscale{1.17}
\plotone{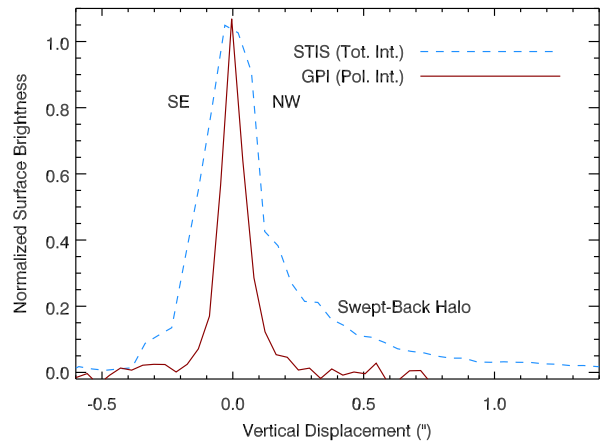}
\caption{Surface brightness profile measured perpendicular to the disk midplane at a distance of 0\farcs75 from the central star (the two sides are averaged). The dashed blue and solid red curves represent the HST/STIS total intensity optical and the GPI $H$-band $Q_\phi$ images, respectively.}\label{fig:halo_profile}
\end{figure}

Outside of the parent body ring, the \hd\ system is characterized by a large-scale halo structure that lies mostly outside of our field-of view. Consistent with past imaging, our observations confirm that the disk extends radially beyond the ring ansae, smoothly connecting the parent body ring and the outer halo. This confirms that the dust located in the halo most likely represents small dust grains that originated in the parent body belt before being radiatively pushed on high eccentricity orbits, where another mechanism then sweeps it out in the NW direction. The fact that the halo is undetected to the NW of the star in our total intensity image, despite this region being the brightest of the halo \citep[e.g.,][]{schneider2014probing} could simply be due to the use of PSF subtraction techniques that effectively cancel out extended, low-gradient surface density structures. On the other hand, our polarized intensity image is free of such effect and, yet, we find no evidence of the presence of the halo. To assess the meaningfulness of this non-detection, we compare our $Q_\phi$ image with the HST/STIS broadband image from \cite{schneider2014probing} in the following manner: we compute surface brightness profiles in 0\farcs15 bands orthogonal to the disk midplane and located 0\farcs75 on either side of the star, i.e., roughly at the disk ansae. Both profiles are averaged to improve signal-to-noise given the lack of marked asymmetry in the halo within the central arcsec. We further rebin the GPI data to roughly match the 0\farcs05 pixel scale of the STIS image. The resulting surface brightness profiles are shown in Figure\,\ref{fig:halo_profile}. The swept-back halo is clearly visible in the STIS surface bright profile, most prominently as an extended structure to the NW of the disk, but is absent in the GPI polarized intensity image with a high degree of significance. Besides the NW extension of the profile, we also find the GPI surface brightness profile to be much narrower around the disk spine than the STIS one. This indicates that the halo also extends radially in front of the parent body ring. 

There are two main possible explanations for the lack of detection of the halo in the STIS $Q_\phi$ image despite the high signal-to-noise detection of the ring itself: either the halo is much bluer than the main ring, or it is characterized by a significantly lower polarization fraction. The latter is inconsistent with the observations that the polarization fraction keeps rising outside of the parent body ring (Figure\,\ref{fig:polar_dist}). On the other hand, the large-scale structure of the system has long been known to be much bluer than the star itself \citep{kalas2005first} whereas the main ring itself is neutral or slightly red \citep[e.g.,][]{esposito2013modeling, rodigas2014does}. We therefore believe that the blue color of the halo is primarily responsible for the lack of detection in our dataset. We also note that the halo is also not apparent to the NW of the star in the $J$-band $Q_\phi$ SPHERE image of the system \citep{bhowmik19}. Overall, the only evidence for the halo in near-infrared images of the system is the curved extension of the disk midplane beyond the ansae due to limb brightening.

\subsection{Scattering properties}

One of the motivations to obtain high fidelity scattered light images of debris disks is to constrain the properties of the dust grains they contain. The surface brightness and color of debris disks are the primary observables affected by dust composition in scattered light images. In addition, polarization measurements provide further information regarding the porosity of grains, since, all else equal, large, porous grains have similar properties to smaller, compact grains \citep{graham2007signature, shen2009modeling}. Assuming that Mie theory accurately describes the scattering properties of dust grains, the minimum grain size and the power law index for the grain size distribution should be tightly constrained by measurements of the scattering phase function and polarization fraction. Indeed, our modeling successfully reproduced both the SPF and the polarizability curve observed for \hd. The size distribution inferred from our modeling, with a minimum grain size of $\approx 1\,\mu$m that is commensurable with the blowout size and a slope consistent with collisional cascade models, is in good agreement both with past studies of the system \citep[including through thermal emission; see for instance][]{donaldson2013modeling} and with general theoretical expectations. 

\begin{figure}
\epsscale{1.17}
\plotone{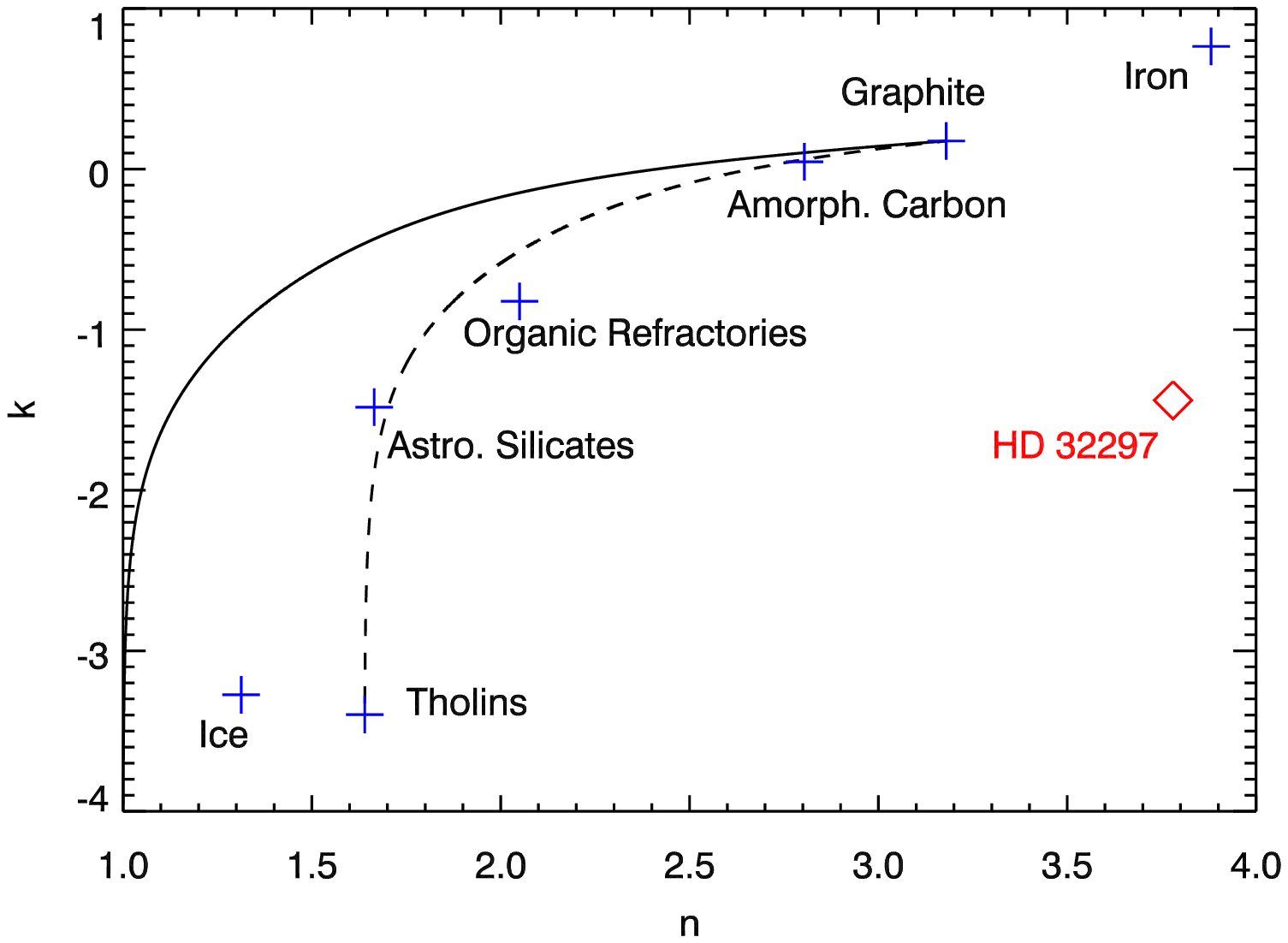}
\caption{Real and imaginary refractive indices at 1.65\,$\mu$m of standard dust species \citep[blue crosses,][]{khare84, draine84, draine85, pollack94, zubko96, li97, li98} and of our best-fitting model to the \hd\ SPF and polarizability curve (red diamond). The solid and dashed curve illustrate the effect of porosity and mixed composition, respectively, using the Bruggeman rule of effective medium theory.}\label{fig:refrac_indices_comparison}
\end{figure}

Despite these apparent successes, it is important to emphasize that the refractive index derived from our analysis lies in a region of the parameter space that is far from all standard dust species, as illustrated in  Figure\,\ref{fig:refrac_indices_comparison}. Worse still, no combination of such species (including void to represent porosity) is consistent with the inferred refractive index. This casts serious doubt on the physical meaning of the other dust parameters that were considered in this analysis. In other words, while we did find a combination of parameters that reproduces well the observed SPF and polarizability curve, it may be the case that this is only a practical empirical model but not one to be trusted at the physical level. There is increasing evidence that dust grains in both the Solar System and in debris disks are aggregates of smaller, sub-micron monomers \citep[e.g.,][]{bentley16}, in which case the Mie model is irrelevant. Unfortunately, despite significant strides towards characterizing the scattering properties of aggregates, it remains beyond the reach of current models to consider aggregates whose size exceed the blow-out size by one or more order of magnitude \citep{arnold19}, which we know are present in debris disks. Furthermore, it may also be instructive to revisit the assumption that the grain size distribution follows a simple power law. Collisional models suggest a more complex underlying structure when factoring in the effects of stellar gravity and radiation pressure in addition to the collisional cascade replenishing the disk \citep[e.g.][]{krivov2006dust, thebault2014grain}.

Leaving aside the physical interpretation of the SPF and polarizability curves, our data provide a robust empirical characterization of the scattering properties of the \hd\ dust ring. With the number of debris disks with estimated SPFs and/or polarizability curves slowly rising, it is now possible to perform model-independent comparisons between systems to identify commonalities and differences between systems. \cite{hughes2018debris} pointed out that most Solar System dust populations share a similar SPF and that the few debris disks with estimated SPFs also match that template. The SPF we have derived for \hd\ is also in reasonable agreement with that ``generic" SPF. On the other hand, the SPF determined by \cite{milli17} for the HR\,4796 ring is markedly different. Combined with the unusual polarization fraction curve observed in that system \citep{perrin15}, this suggests that this latter disk is characterized by a markedly different dust population. While such comparisons are best performed by extracting the SPF from observations, this process suffers from possible ambiguities and possible biases, as we have already discussed. 

\begin{figure}
\epsscale{1.17}
\plotone{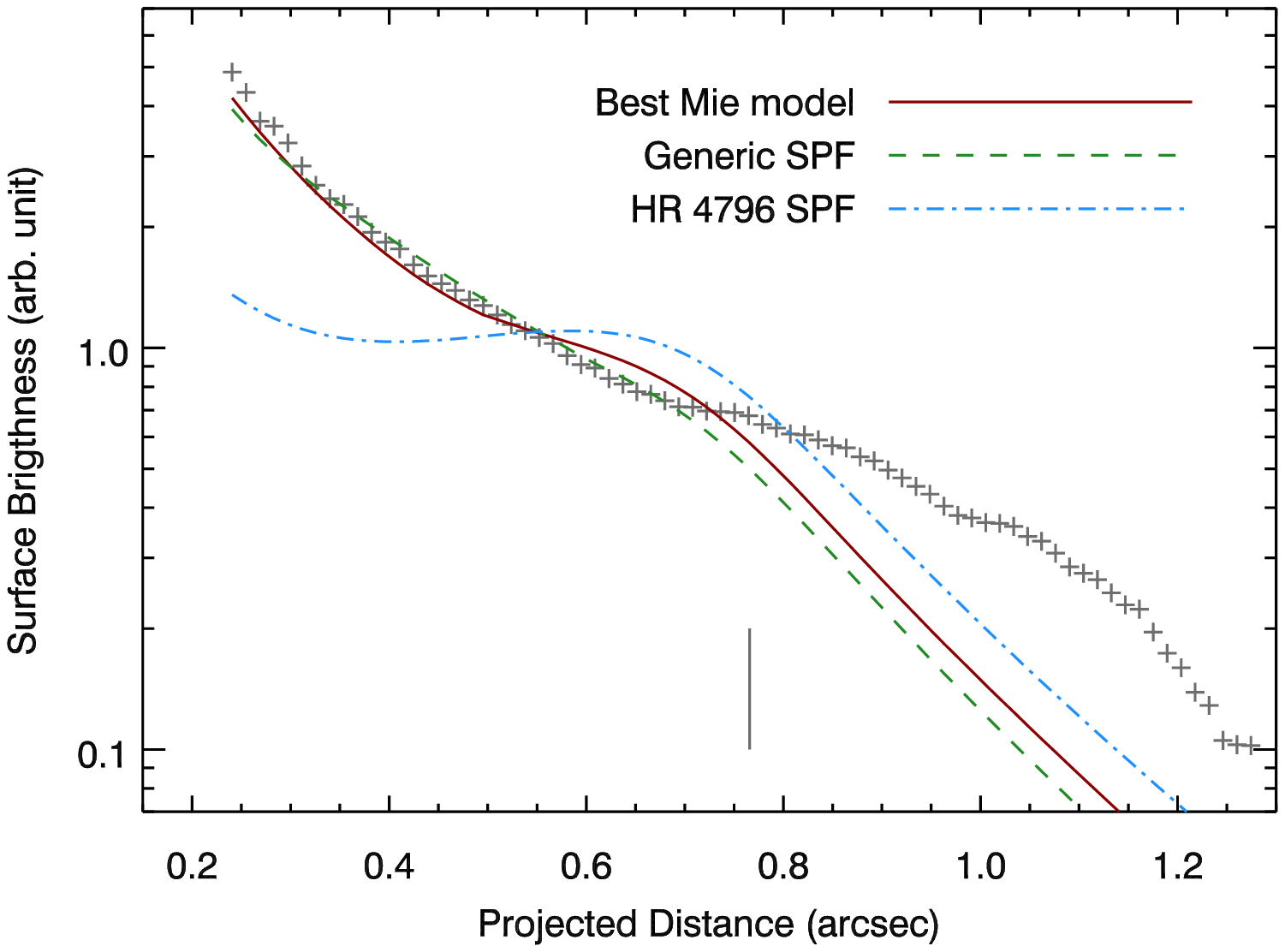}
\caption{Observed $H$ band total intensity surface brightness profile for \hd\ (gray plus signs) compared to predicted profiles assuming different SPFs. The solid red, dashed green and blue dot-dashed curves correspond to the best-fitting Mie model derived in this study, the generic SPF identified by \cite{hughes2018debris} and the HR\,4796 SPF from \cite{milli17}, respectively. The vertical segment marks the location of the ring radius, as inferred from our geometric analysis (\S\,\ref{subsec:morph}).}\label{fig:spf_compar}
\end{figure}

To illustrate the effects of a different SPF on the appearance of a debris disk, we compared the modeled surface brightness profile along the spine of a nearly edge-on disk (using the geometric parameters indicated in Table\,\ref{tab:geom_fit}) assuming three distinct SPFs: the best fitting Mie model presented in Section\,\ref{subsec:dustprops}, the ``generic" SPF from \cite{hughes2018debris} and the HR\,4796 SPF from \cite{milli17}. The latter SPF is not defined at all scattering angles due to our particular viewing geometry of the system, so we performed linear extrapolations the SPF for scattering angles $<15$\degr and $>165$\degr. These regions are behind the coronagraphic mask once the disk is observed with the viewing geometry of \hd, therefore the details of this extrapolation are not critical to the comparison. The results of this exercise are illustrated in Figure\,\ref{fig:spf_compar}. All three models under-predict the surface brightness profile outside of the main ring radius, but both the best Mie model and the generic SPF match the data extremely well inside of that projected distance. This confirms that the total intensity scattering properties of the \hd\ dust is consistent with most other astrophysical dust populations. Conversely, if this disk was characterized by an SPF that is similar to that observed for HR\,4796, its surface brightness profile would be dramatically different, with a nearly flat surface brightness profile that is inconsistent with the observations. This is due to the combination of 1) the fact that the HR\,4796 SPF has its minimum at a scattering angle of $\approx50$\degr\ with significant backscattering at angles $\gtrsim100$\degr, and 2) limb brightening in the optically thin ring. This is further evidence that the scattering properties in the \hd\ and HR\,4796 debris disks are clearly distinct.

An additional qualitative feature of the SPF in \hd\ is the sharp peak observed in polarized intensity close to the inner working angle of our observations. Since the polarization fraction is expected by symmetry to drop to zero at 0\degr\ scattering angle, this indicates the SPF itself must be characterized by a very sharp forward scattering peak, reminiscent of the HR\,4796 and $\beta$\,Pic \citep[][Arriaga et al. in prep]{perrin15, mmb15}. On the other hand, there are several edge-on debris disks that have been imaged in polarized intensity and that do not show such a feature \citep[][Esposito et al., submitted]{olofsson16, engler17, esposito18}. This further hints at the fact that the scattering properties of dust populations in debris disks are not all identical, although interpreting them in terms of physical properties of the grains may still be out of reach.

Finally, another qualitative approach to constraining the SPF in the case of nearly edge-on disks like \hd\ is to assess whether the back side of the ring contributes to the scattered light image. \cite{bhowmik19} presented a tentative detection in total intensity using ADI PSF subtraction, which would imply a strong backscattering peak. We do not confirm this feature in our observations of \hd. It is possible that differences between PSF subtraction introduce different artefacts, precluding a definitive conclusion. However, we point out that, given the derived disk radius and inclination, the projected separation along the minor axis between the front and back side of the ring is in the 0\farcs04-0\farcs07 range, which makes it extremely challenging to detect. In polarized intensity, which is unaffected by PSF subtraction, the lack of an increase in vertical extent of the disk just inside of the ansae indicates that the back side of the ring does not contribute significantly to the observed surface brightness, indicating that the SPF drops significantly around a scattering angle of 90\degr. This latter conclusion matches our conclusion that the SPF of \hd\ is qualitatively different from that observed in HR\,4796. 

\subsection{Underlying planetary system}

Debris disks are believed to be associated with planetary-mass objects, as readily illustrated in some well-known systems, such as $\beta$\,Pic and HR\,8799. No point source is evident in our total intensity image of the system. We note, however, that observations with GPI's polarization mode are not optimal for detection of planets unless they are highly linearly polarized. Instead, integral field spectroscopy observations provide the deepest search for planets.\cite{bhowmik19} presented such observations of \hd, reaching a contrast limit of $10^{-5}$ or better outside of 0\farcs5. At an assumed age of 30\,Myr, this corresponds to an upper limit on any planetary mass object of 4--5\,$M_\mathrm{Jup}$ based on the COND models \citep{baraffe2003evolutionary}. An important caveat, however, is that the detection limit is much worse along the bright disk spine, so that the upper limit quoted computed only applies outside of the plane of the disk. Specifically, a $10^{-5}$ contrast point source would have its peak pixel brightness equal that of the disk, and therefore would be marginally detectable at best, at a distance of $\approx1\farcs2$ from the star if it lies in the plane of the disk. At closer separation, the contrast degrades proportionally to the disk peak surface brightness, reaching $10^{-4}$ (or a planet mass of $\approx12\,M_\mathrm{Jup}$) at $\approx0\farcs45$.

An indirect probe of the presence of planetary mass bodies in the system is through their dynamical interaction with the disk. The lack of significant lateral asymmetry and of local (photometric or morphological) perturbation in the parent body belt supports the picture of a dynamically cold, azimuthaly symmetric system, seemingly ruling out strong planet-disk interactions. The scale height of the belt can be also related to its dynamical excitation, since the scale height is directly related to the velocity dispersion of the solid bodies. The disk scale height derived from our geometric analysis ($h/r \approx 0.04$) is consistent with dynamical models that only consider collisions between grains and radiative forces \citep{thebault2009vertical}. Therefore, the dynamical state of the \hd\ main belt can be fully explained without invoking the presence of planetary-mass bodies stirring the system. A planet could however be responsible for the inner dust depletion (inside of 30--50\,au) without introducing measurable local perturbation. In addition, \cite{lee2016} proposed that an interior planet on an inclined orbit is responsible for the "double wing" in the extended outer halo \citep{schneider2014probing}, although this could also arise from interaction with secondary gas \citep{lin19}. Current observations of the \hd\ system are thus inconclusive regarding the presence and structure of its planetary system.

\section{Conclusion} \label{sec:conclu}

As part of the commissioning phase of the GPI instrument, we have obtained $H$ band high-contrast total and polarized intensity images of the edge-on \hd\ debris disk. The disk is detected from just outside the edge of the coronagraphic mask, $\approx0\farcs15$ from the central star, out to edge of the field-of-view, at projected distance of $\approx1\farcs3$. Using the slight curvature of the disk spine, we determined the disk geometry and found that the disk radius is $\approx100$\,au, smaller than previous scattered light studies of the system, highlighting the difficulty of measuring disk size in an edge-on configuration. However, since the radius we derive is consistent with the thermal emission images of the disk, we believe that it represents the true size of the parent body belt. 

We applied four multiple PSF subtraction post-processing algorithms and demonstrated that three of these methods yield reliable surface brightness distributions in the case of an edge-on disk. Using these, we found that the disk is consistent with being azimuthally symmetric. We also estimated the SPF and polarizability curves of the dust present in the disk. We find curves that are typical of Solar System dust populations and of other debris disks, with the marked exception of the HR\,4796 debris disk. Assuming Mie scattering, we find a dust model that simultaneously reproduces the SPF and polarizability curves, but the resulting refractive index is inconsistent with any standard dust composition. The most likely explanation is that dust grains in the system are not compact spheres but complex aggregates, as seen in Solar System dust populations. 

The large scale swept-back halo present outside the parent body ring is undetected in our data, confirming that it contains primarily sub-micron grains that produce blue scattering. Finally, we do not detect any planetary-mass object in the system, although we stress that our detection limit is severely limited by the bright disk for objects whose orbit is coplanar with the disk itself. Given the symmetric and small vertical extent of the parent body belt, we find no evidence for stirring induced by an unseen planetary-mass body although we cannot exclude the presence of an object that is sufficiently distant from the belt.

\acknowledgments{We are grateful to Glenn Schneider for making his STIS image available for analysis. This work is based on observations obtained at the Gemini Observatory, which is operated by the Association of Universities for Research in Astronomy, Inc. (AURA), under a cooperative agreement with the National Science Foundation (NSF) on behalf of the Gemini partnership: the NSF (United States), the National Research Council (Canada), CONICYT (Chile), Ministerio de Ciencia, Tecnolog\'ia e Innovaci\'on Productiva (Argentina), and Minist\'erio da Ci\^encia, Tecnologia e Inova\c c\~ao (Brazil). This work made use of data from the European Space Agency mission {\it Gaia} (\url{https://www.cosmos.esa.int/gaia}), processed by the {\it Gaia} Data Processing and Analysis Consortium (DPAC, \url{https://www.cosmos.esa.int/web/gaia/dpac/consortium}). Funding for the DPAC has been provided by national institutions, in particular the institutions participating in the {\it Gaia} Multilateral Agreement. This research made use of the SIMBAD and VizieR databases, operated at CDS, Strasbourg, France.

Supported by NSF grants AST-1411868 (E.L.N., K.B.F., B.M., and J.P.), AST-141378 (G.D.), and AST-1518332 (T.M.E., R.J.D.R., J.R.G., P.K., G.D.). Supported by NASA grants NNX14AJ80G (E.L.N., B.M., F.M., and M.P.), NNX15AC89G and NNX15AD95G/NExSS (T.M.E., B.M., R.J.D.R., G.D., J.J.W, J.R.G., P.K.), NN15AB52l (D.S.), and NNX16AD44G (K.M.M.). M.R. is supported by the NSF Graduate Research Fellowship Program under grant number DGE-1752134. J.R. and R.~Doyon acknowledge support from the Fonds de Recherche du Qu\`ebec. J.~Mazoyer's work was performed in part under contract with the California Institute of Technology/Jet Propulsion Laboratory funded by NASA through the Sagan Fellowship Program executed by the NASA Exoplanet Science Institute. M.M.B. and J.~Mazoyer were supported by NASA through Hubble Fellowship grants \#51378.01-A and HST-HF2-51414.001, respectively, and I.C. through Hubble Fellowship grant HST-HF2-51405.001-A, awarded by the Space Telescope Science Institute, which is operated by AURA, for NASA, under contract NAS5-26555. K.W.D. is supported by an NRAO Student Observing Support Award SOSPA3-007. J.J.W. is supported by the Heising-Simons Foundation 51 Pegasi b postdoctoral fellowship. This work benefited from NASA's Nexus for Exoplanet System Science (NExSS) research coordination network sponsored by NASA's Science Mission Directorate. Portions of this work were also performed under the auspices of the U.S. Department of Energy by Lawrence Livermore National Laboratory under Contract DE-AC52-07NA27344.}

\software{{\tt MCFOST} (\citealt{pinte2006monte}), Gemini Planet Imager Data Pipeline (\citealt{perrin14, perrin2016_drp}, \url{http://ascl.net/1411.018}), pyKLIP (\citealt{wang15pyklip}, \url{http://ascl.net/1506.001}), NumPy (\citealt{numpy}, \url{https://numpy.org}), SciPy (\citealt{scipy}, \url{http://www.scipy.org/}), Astropy \citep{astropy2018}, matplotlib \citep{matplotlib2007, matplotlib_v2.0.2}, iPython \citep{ipython2007}, emcee (\citealt{foremanmackey2013emcee}, \url{http://ascl.net/1303.002}), corner (\citealt{corner}, \url{http://ascl.net/1702.002})}.

\facilities{Gemini:South}

\clearpage

\appendix
\section{Surface brightness preservation of PSF subtraction methods: Injection and recovery test} \label{subsec:recovery}

The various PSF subtraction methods we employed suffer from several potential limitations: self- and over-subtraction, significant correlated residuals, poor sensitivity to smooth, extended surface brightness such as the \hd\ halo, to name the most important ones. Besides assessing the geometry of the disk, our goal is to measure the surface brightness profile of the disk, which requires understanding the amplitude of these effects. Forward modeling, which maps out the throughput of PSF subtraction methods, is necessary when applying the standard ADI method \citep[e.g.,][in the case of \hd]{boccaletti2012morphology,esposito2013modeling}, but ultimately the precision of the process is limited by the fact that 1) the self- and/or over-subtraction is a large fraction of the input surface brightness, and 2) a bright disk like in the case of \hd\ results in a breakdown of the assumption that the astrophysical signal is small compared to the stellar PSF brightness. 

To circumvent the limitations of ADI, we used the RDI, MI and NMF methods, which we expect to result in small (or negligible) surface brightness loss, at least along the disk spine. To assess the reliability of these methods, we perform an injection-and-recovery test. Out of all the reference frames used in the RDI subtraction, we identified the HIP\,46634 dataset (taken on 2015 February 1) as the most correlated with the \hd\ one. Visual inspection confirmed the clear similarities in the PSF structure between both datasets. We generated a synthetic disk model. For simplicity, we assumed the disk to be exactly edge-on, a surface brightness profile that obeys follows an $r^{-1.5}$ power law, and a vertical structure characterized by a $\approx2$-pixel FWHM Gaussian profile. Those choices were made to roughly match the appearance of the \hd\ disk. The disk model image was then convolved by a 4-pixel FWHM two-dimensional Gaussian appropriate for GPI in the $H$ band and the disk surface brightness was scaled relative to the input dataset to match the observed disk. In particular, since HIP\,46634 is about 0.8\,mag brighter in $H$-band than \hd, the surface brightness of the injected model is higher than that of the \hd\ disk. We then injected this disk in each individual frame of HIP\,46634 and performed the same PSF subtraction process as described above. We did not use ADI in this test, as it has already been established that this method does not preserve flux of extended disks \citep[][in the case of \hd]{esposito2013modeling}.

\begin{figure}
\plotone{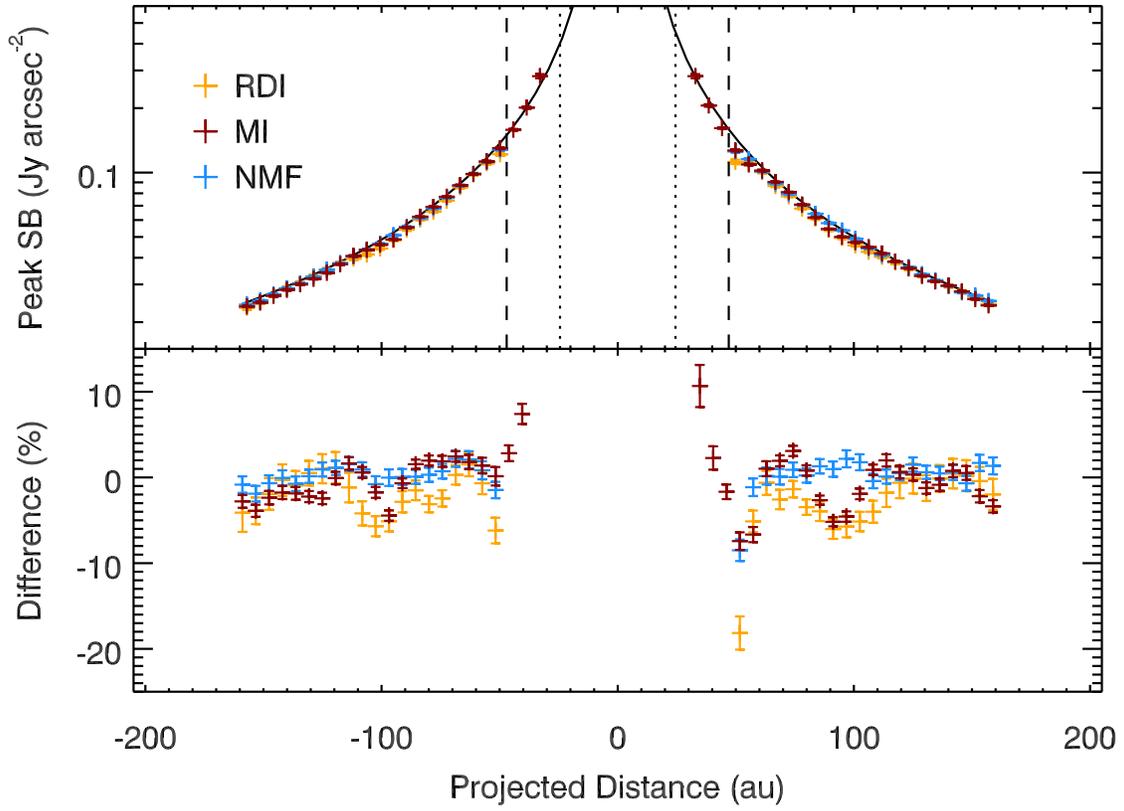}
\caption{{\it Top:} Input (solid black curve) and retrieved peak surface brightness of a model edge-on disk injected in a disk-free GPI dataset. The results from three PSF subtraction methods are shown here: RDI, MI and NMF (orange, blue and red symbols, respectively). The vertical dotted (dashed) lines indicate the smallest separations at which the MI- (RDI- or NMF-)processed image can be trusted. {\it Bottom:} Relative difference between the input and output surface brightness profiles. \label{fig:disk_retrieval}}
\end{figure}

Figure\,\ref{fig:disk_retrieval} shows the resulting inferred surface brightness profiles compared to the input model. The RDI, MI and NMF methods retrieve the full surface brightness of the disk's spine to within 10\%, albeit with some radial substructures. We stress that the degree of reliability may depend on the specific dataset used in the injection. Nonetheless, it is encouraging that peak surface brightness is preserved to such high precision by the different methods we employed. On the other hand, measuring integrated brightness in vertically extended boxes centered on the disk result in larger discrepancies (up to $\approx 30\%$), as the PSF subtraction process tends to remove some signal in the lower surface brightness regions away from the disk spine. Despite these imperfections, this analysis confirms that all three non-ADI PSF subtraction methods preserve the total intensity profile of the disk along its spine.




\clearpage 

\bibliography{Bibliography}{}
\bibliographystyle{aasjournal}

\end{document}